\documentclass[aps,prd,onecolumn,groupedaddress,floatfix,longbibliography,superscriptaddress,notitlepage,keywords]{revtex4-2}

\usepackage{soul}
\usepackage{epsfig}
\usepackage{subfigure}
\usepackage{dcolumn}
\usepackage{bm}
\usepackage[usenames,dvipsnames]{xcolor}
\usepackage{slashed}
\usepackage{graphicx,color}

\usepackage{appendix}
\usepackage{tabularx}
\usepackage{amssymb}
\usepackage{amsmath,latexsym}

\usepackage[bookmarksnumbered, pdfstartview=FitH,colorlinks,urlcolor=blue, citecolor=blue,linkcolor=blue] {hyperref}

\usepackage{ulem}

\newcommand{\beq}{\begin{equation}}

\newcommand{\eeq}{\end{equation}}

\newcommand{\mbf}{\mathbf}

\begin{document}

\preprint{APS/123-QED}

\date{\today}

\title{A supersymmetric study of charge and spin transport in Weyl semimetals under axionic electrodynamic response}

\author{J. C. Pérez-Pedraza\footnote{Corresponding author}}
\email{julio@correo.nucleares.unam.mx}
\affiliation{Instituto de Ciencias Nucleares, Universidad Nacional Aut\'{o}noma de M\'{e}xico, 04510 Ciudad de M\'{e}xico, M\'{e}xico}

\author{A. Martín-Ruiz}
\email{alberto.martin@nucleares.unam.mx}
\affiliation{Instituto de Ciencias Nucleares, Universidad Nacional Aut\'{o}noma de M\'{e}xico, 04510 Ciudad de M\'{e}xico, M\'{e}xico}

\author{L. F. Urrutia}
\email{urrutia@nucleares.unam.mx}
\affiliation{Instituto de Ciencias Nucleares, Universidad Nacional Aut\'{o}noma de M\'{e}xico, 04510 Ciudad de M\'{e}xico, M\'{e}xico}

\begin{abstract}
We investigate the charge and spin transport properties of a Weyl semimetal under an external magnetic field using a low-energy effective theory. By performing a chiral transformation, we remove the axial term from the fermionic Lagrangian, while its physical effects are retained through the system's electromagnetic response. This response, derived from integrating out the fermionic degrees of freedom, takes the form of axionic electrodynamics and enters the Dirac equation via minimal coupling. The resulting dynamics separate naturally into magnetic and electric components: the magnetic sector admits a supersymmetric factorization, while the electric part exhibits a PT-supersymmetric structure. { Robin  boundary conditions are applied to the spinor, setting the energy spectrum and defining exact spinor solutions. We obtain the chiral projections of probability and current densities, exhibiting a chiral imbalance in the material. We show that in the $x$-direction, only one chirality contributes to the chiral current ($j_l^x=0$). Spin density and currents where also computed.} Our results demonstrate how the interplay between axionic response and supersymmetry governs transport phenomena in our specific setup, offering novel insights into the effective field theory description of Weyl semimetals.\\

\end{abstract}

\maketitle

\section{Introduction}

The theoretical and experimental discovery of topological insulators \cite{Hasan_RevModPhys.82.3045,Qi_RevModPhys.83.1057} triggered a paradigm shift in condensed matter physics, establishing topological phases as a central topic of study. Within this landscape, topological semimetals have emerged as a distinct and intriguing class, with Weyl semimetals (WSMs) representing a prominent example \cite{burkov2016topological,armitage_RevModPhys.90.015001,yan_annurev:/content/journals/10.1146/annurev-conmatphys-031016-025458}. WSMs are characterized by linear band-touching points in the electronic spectrum (Weyl nodes) where conduction and valence bands intersect near the Fermi energy. These nodes appear in pairs of opposite chirality (guaranteed by the Nielsen-Ninomiya theorem \cite{NIELSEN1983389}) and act as monopoles of Berry curvature in momentum space, endowing the system with robust topological properties protected by a nonzero Berry flux through the Fermi surface. The bulk-boundary correspondence predicts the emergence of Fermi arcs, open surface states in the surface Brillouin zone, which connect the projections of Weyl nodes with opposite chirality \cite{wan_PhysRevB.83.205101}.

The effective Hamiltonian near the Weyl nodes can be expressed as:
\begin{align}
    H = \hbar v _{F} \gamma ^{0} \gamma ^{\mu} \left( k _{\mu} - b _{\mu} \gamma ^{5} \right)  ,
    \label{BASICHAM} 
\end{align}
where $v _{F}$ is the Fermi velocity characteristic of the semimetal, $k _{\mu} = (\omega / v _{F} , \mathbf{k})$ is the crystal momentum operator, $b _{\mu} = (b _{0} / v _{F} , \mathbf{b})$ is a background axial 4-vector that encodes the separation of the Weyl nodes, $\gamma ^{\mu}$ are the Dirac gamma matrices in 3+1 dimensions and $\gamma ^{5} = i \gamma ^{0} \gamma ^{1} \gamma ^{2} \gamma ^{3}$ is the chirality matrix. The four position is given by $x ^{\mu} = ( v _{F} t, \mathbf{x})$.

The presence of the axial term $b _{\mu} \gamma ^{5}$ explicitly breaks inversion (I) and time-reversal (TR) symmetries and distinguishes the two Weyl nodes by their chirality. This term is responsible for various anomalous transport phenomena and topological responses characteristic of Weyl semimetals.

It is important to emphasize that the low-energy effective model (\ref{BASICHAM}) is valid only in the vicinity of the Weyl nodes, where the linear approximation of the band structure holds. The effects arising from lattice regularization, band curvature, and higher energy bands are neglected. Thus, phenomena occurring at energy scales far from the node or involving intervalley scattering may require more comprehensive microscopic models.

At low energies, quasiparticles near each node behave as relativistic Weyl fermions, described by an effective Dirac equation modified by an axial coupling
 \beq
 (i \hbar v_F \gamma^\mu \partial_\mu-e\gamma^\mu A_\mu +(\partial_\mu \theta) \gamma^\mu \gamma^5) \psi=0, \qquad {b_\mu= \frac{1}{2}\partial_\mu \theta.}
 \label{DIRAC0}
\eeq
{
This axial term encodes the separation in momentum and/or energy between Weyl nodes. More precisely, the spatial component $\mathbf b$ characterizes the momentum-space separation of the nodes, whereas the temporal component $b_0$ accounts for a possible energy offset between nodes of opposite chirality. Such separations arise from the breaking of either parity  \cite{huang2015theoretical,weng_PhysRevX.5.011029,xu2015discovery,lu2015experimental,lv2015experimental,lv2015observation,xu2015experimental,lv2015observation1,liu2016evolution}, time-reversal \cite{witczak_PhysRevB.85.045124,chen_PhysRevB.86.235129,gao2021time}, or both symmetries simultaneously \cite{zyuzin2012weyl}, and have been experimentally observed in several Weyl semimetals, including TaAs and related compounds, through ARPES and transport measurements~\cite{Xu2015,Lv2015,Yang2015,Xu2015Science}.

}

In this work we aim to explore the one-particle effects of this axial coupling in describing the fermionic response of a WSM under prescribed electric and magnetic fields. Naively, one would think that Eq. (\ref{DIRAC0}) is the suitable starting point for solving the Dirac field in terms of external electromagnetic fields. However, this is not possible since the chiral transformation
\begin{align}
    \Psi  (x) \quad \to \quad  \Psi ' (x) =e ^{i \theta (x)  \gamma ^{5} / 2 } \, \Psi  (x) , \qquad \theta (x) = 2 x ^{\mu} b _{\mu} ,
    \label{CHIRALT}
\end{align}
allows us to eliminate the axial vector term $b _{\mu} \gamma ^{5}$ from the Dirac operator \cite{zyuzin}. {Typically, the parameter  $\theta$ includes discontinuous boundary terms that account for nearby materials. These terms are fully eliminated by the transformation (\ref{CHIRALT}),  after which  the effective} Dirac equation simplifies to
\begin{align}
\label{2}
    \left( i \hbar v _{F} \gamma ^{\mu} \partial _{\mu} - e \gamma ^{\mu} A _{\mu} \right) \, \Psi ' (x) = 0 ,
\end{align}
where $A _{\mu}$ is the electromagnetic gauge field, which will encode the axionic response. While this transformation removes the axial coupling explicitly from the fermionic action, the underlying physics associated with Weyl node separation is preserved and manifests in the electromagnetic response. In other words, we expect the electromagnetic fields to satisfy modified Maxwell's equations including the Weyl nodes effects. This modification arises from the calculation of the effective electromagnetic action in quantum field theory and is consistent with the well-known fact that when integrating the fermionic fields in the corresponding path integral expression, it is necessary to take into account the Jacobian of the axial fermionic transformation since the path integral measure is not invariant under this transformation. Formally, the fermionic partition function before and after the chiral transformation relates as
\begin{align}
    \mathcal{Z} [A _{\mu}] = \int d \Psi d \bar{\Psi} \, e ^{ \frac{i}{\hbar} S [\Psi , \bar{\Psi}, A _{\mu} ] } = \int d \Psi ' d \bar{\Psi} ' \, J [\theta (x)] \, e ^{ \frac{i}{\hbar} S [\Psi ' , \bar{\Psi} ', A _{\mu} ] }.
\end{align}
Here $J [\theta (x)]$ is the Jacobian factor generated by the chiral rotation, yielding an additional term in the effective electromagnetic action, which takes the form of an axion coupling:
\begin{align}
    S _{\theta} = \frac{e ^{2}}{2 \pi h} \int  \theta (x) \, \mathbf{E} \cdot \mathbf{B} \, d ^{4} x , 
\end{align}
where $\theta (x) = 2 b _{\mu} x ^{\mu}  $ acts as an axion-like field proportional to the Weyl node separation, and $\mathbf{E} $ and $\mathbf{B}$ are the electric and magnetic fields, respectively. The contribution $S_\theta$ added to the standard Maxwell's action defines the theory known as axion electrodynamics \cite{Sikivie:1983ip}, a specific case of which is Carroll-Field-Jackiw electrodynamics \cite{Carroll:1989vb}. 

Therefore, the low-energy physics of Weyl semimetals in the presence of an external electromagnetic field can be equivalently described by a Dirac fermion without explicit axial coupling but subject to an effective electromagnetic response governed by axionic electrodynamics.

We choose to begin with a constant magnetic field and take the electric field as that induced by the magnetoelectric effect derived from axion electrodynamics, to first order in $\theta$. We couple these two fields to the Dirac equation via the minimal substitution and study the consequences of such effect upon the fermionic field $\Psi$ which solves the quantum mechanical Dirac equation with no axial coupling. In other words, we look for the fermionic response of the WSM sourced by axion ED, i.e., we start from a standard Dirac equation coupled to electromagnetic fields $A_\mu$ which descend from axion electrodynamics. This Dirac equation (\ref{2}) keeps its invariance under global axial transformations such that the corresponding axial current ${\bar \Psi}\gamma^\mu \gamma^5 \Psi$ is still conserved. In other words, and because we remain within an effective one-particle description, no anomalous divergence of the axial current arises at this stage. {This does not imply the absence of anomaly-related physics. Rather, the effects of the chiral anomaly have already been incorporated through the axion-electrodynamics term generated by the Jacobian of the chiral transformation, whose associated electromagnetic fields provide the background in which the fermionic dynamics are studied.} The conservation of the axial current will therefore serve as a consistency check of our results.

Motivated by previous work in Ref. \cite{perez2024dirac} we {conveniently} choose the directions of the EM fields {and find} that the resulting quantum mechanical system exhibits not only standard supersymmetry but also PT supersymmetry. Both mechanisms follow into the broader factorization methods developed in Ref. \cite{infeld}. Perhaps the most well known of these realizations is encapsulated in supersymmetric quantum mechanics, which  can be viewed as a supersymmetric field theory in one time and zero spatial dimensions. In standard quantum field theory,  supersymmetry (SUSY) arised as a mechanism to override and extend the Coleman Mandula theorem \cite{CM}, which states that is not possible to  mix internal symmetries with Poincaré symmetry. The extension was made possible by replacing the reliance on Lie algebras with Lie superalgebras to quantify the generators of the symmetries. SUSY appeared as a promising way to attain a finite theory of gravity via its generalization to supergravity. However the non-renormalizability of  $N=1$ supergravity gravity reappears at the two-loop order and  higher. Also  SUSY  was found to be useful in canceling  divergences in the Standard Model of particles. The basic idea is to mix fermions and bosons in specific multiplets transforming covariantly under the Poincaré superalgebra. Nevertheless, this required the introduction of a plethora of additional particles beyond those encountered in the Standard Model, leading to its diverse supersymmetric extensions. The relevance of SUSY in high energy physics is still under scrutiny, but up to day no supersymmetric partners of the known elementary particles have been found. For a review of the vast literature in this subject see for example \cite{feng,REVSUSY1,REVSUSY2}.  
However, the introduction of these ideas has been very significant from a theoretical point of view in many areas of physics. In particular, SUSY  in quantum mechanics of various dimensions generates a wide variety of exactly solvable systems, providing useful relationships between the spectra and the dispersion parameters that appear in the different channels of the model. The original supersymmetric quantum mechanics (SUSY-QM) was formulated in terms of hermitian models \cite{WITTEN}, but it  was subsequently extended to the class of  non-hermitian system characterized for respecting PT symmetry \cite{BENDER}. 
Recently there has been some interest in studying the applications of SUSY to  Dirac materials under the presence of given external electric and magnetic  fields in the quest of incorporating the corresponding axion electrodynamics that describes their  electromagnetic interaction. We find applications in graphene \cite{kuru,concha, jakubsky, ioffe, schulze, bagchi}  and in the planar Hall effect \cite{perez2024dirac}, {where the Hall response occurs in the plane spanned by the electric and magnetic fields, rather than perpendicular to it as in the conventional Hall effect.}

In the subsequent sections we build upon  the framework outlined in the Introduction, incorporating the axionic electromagnetic response of a WSM  into the axial-free Dirac equation, which we then solve using supersymmetric techniques.

The paper is organized as follows. {In Section \ref{electrodynamics}, we study the electromagnetic response in the interior region of a Weyl semimetal slab of finite thickness subject to an external magnetic field, deriving the associated potentials that will enter the Dirac equation via minimal coupling.
}. Section \ref{Dirac_eq_section} is devoted to the explicit construction of Dirac spinors in the Weyl representation, where we clearly separate the effects of electric and magnetic fields using techniques from supersymmetric quantum mechanics. In Section \ref{currents_section}, we analyze the resulting charge and spin transport properties. {Section \ref{DEG} includes a short discussion on how to determine the quantum numbers that label the degeneracy of the system.} We close with  Section \ref{conclusions}, where we  present our conclusions and outlook. 
{There are seven  Appendices where most of the detailed calculations are presented. The  Appendix \ref{FACTHAM} provides a summary of the key results from the factorization method used throughout the manuscript in the framework of supersymmetric quantum mechanics. In the Appendix \ref{APPB} we recap how current conservation arises directly from the general expression of the  classical Dirac equation. The Robin boundary  conditions required for the interior problem are discussed in the Appendix \ref{NORM2} and their specific implementation is carried out in the Appendix  \ref{APPD}. The final form of the spinor having only one arbitrary  constant to be determined by normalization is obtained in the Appendix \ref{APPE}. The  Appendix \ref{APPF} includes an additional proof of current conservation, starting from the specific expressions  of the currents in our case. Finally Appendix \ref{INTERT} {presents  the calculation of the intertwining relations in the electric sector}.}

\section{Axion Electrodynamics in Weyl Semimetal Slab} \label{electrodynamics}

\begin{figure}
\begin{center}
        \includegraphics[width=0.5\textwidth]{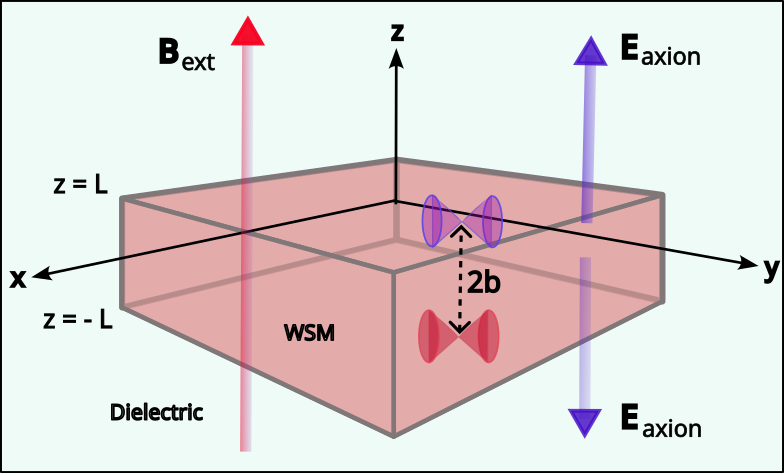}
        
        \caption{System under study consisting of a Weyl-semimetal slab of thickness $2L$ along the $z$-direction immersed in an external magnetic field parallel to the $z$-axis. The axion magnetoelectric response in the material induces a nonhomogeneous electric field in the same direction. We consider the interior problem only, i.e. in the region $-L < z < +L $. }
    \label{system}
\end{center}    
\end{figure}

In this section, we introduce the physical setup under consideration: the bulk of a topological Weyl semimetal (WSM) modeled as a plane slab of finite thickness $2L$ along the $z$-direction, extending infinitely in the transverse directions. The bulk of the material hosts a pair of Weyl nodes located at the neutrality point and separated along the $z$-axis in momentum space \cite{PhysRevB.99.155142}. The region outside the slab is filled with a dielectric medium, ensuring continuity of the electromagnetic fields across the interface. The whole system is subjected to a uniform external magnetic field $\textbf{B} _{0} = B _{0}  \hat{\textbf{e}} _{z}$. A schematic representation of the configuration is shown in Fig.~\ref{system}.

In the absence of free charges and currents, the axion electrodynamics equations take the form:
\begin{align}
\nabla \cdot \mathbf{D} =  \frac{e ^{2}}{2 \pi h} \, \nabla \theta \cdot \mathbf{B} , \quad \nabla \cdot \mathbf{B} = 0 , \quad \nabla \times \mathbf{E} = - \frac{\partial \mathbf{B} }{\partial t} , \quad   \nabla \times \mathbf{H} = - \frac{e ^{2}}{2 \pi h} \nabla \theta\times \mathbf{E} ,
\label{AXIONED}
\end{align}
where $\theta (\mathbf{r}) = 2 \mathbf{b} \cdot \mathbf{r} $ is the axion-like field, and $\mathbf{D} = \epsilon \mathbf{E}$ and $\mathbf{H} = \mathbf{B} / \mu $ are the displacement and magnetizing fields, respectively.  The derivatives of $\theta$ act as effective sources and currents, encoding the topological structure of the Weyl semimetal. These modified Maxwell equations will now be solved for the slab geometry introduced above.  

The system is initially subjected to an external magnetic field $\mathbf{B} _{0}$, which initiates the axionic electromagnetic response. Given that the magnetoelectric coupling is weak (being proportional to the fine-structure constant $\alpha = 1/ (4 \pi \epsilon _{0} \hbar c) \approx 1/137$), we approach this interaction perturbatively. To leading order in $B_{0}$, the modified Maxwell equations can be solved linearly, yielding induced electric and magnetic fields $\mathbf{E}$  and $\mathbf{B}$ that are determined by:
\begin{align}
\nabla \cdot \mathbf{E} =  \frac{e ^{2}}{2 \pi h \epsilon} \, \nabla \theta \cdot \mathbf{B} _{0} , \qquad \nabla \times \mathbf{B} =  - \frac{\mu _{0} e ^{2}}{2 \pi h}  \nabla \theta\times \mathbf{E},  \label{Modified_Eqs}
\end{align}
respectively.  Here, $\epsilon$ is the dielectric constant and $\mu_{0}$ the vacuum permeability.

The interior of the slab configuration in Fig. \ref{system} is captured by the position-dependent axion angle: $\theta (z) =2 b z \, H (L-z) H (L+z)$, such that
\begin{align}
\nabla \theta = 2b \, H (L-z) H (L+z) - 2b L \left[ \delta (L-z) + \delta (L+z)  \right] . 
\end{align}
This leads to a bulk charge density $\rho _{\theta} (z) = \rho _{0}   \, H (L-z) H (L+z) $ together with a surface charge density $\sigma _{\theta} (z) = - \rho _{0}  L \delta (L-z) - \rho _{0}  L  \delta (L+z)  $, where $\rho _{0} = \frac{e ^{2} b }{  \pi h} B _{0} $. The electric field can therefore be obtained from
\begin{align}
\nabla \cdot \mathbf{E} =  \frac{1}{\epsilon} \left[ \rho _{\theta} (z) + \sigma _{\theta} (z) \right] .
\end{align}
Owing to the symmetry of the problem, one can use the Gauss law to compute the electric field. Taking a Gaussian pillbox inside the slab we get
\begin{align}
\mathbf{E} _{i} (z) = \frac{\rho _{0} z}{\epsilon}  \, \hat{\mathbf{e}} _{z} , \qquad  \vert z \vert < L,
\label{E-field}
\end{align}
while for the outside region $\mathbf{E} _0 (z) = 0$, since the net electric charge enclosed by a pillbox outside the slab is zero, i.e. $\int _{-L} ^{L} \left[ \rho _{\theta} (z) + \sigma _{\theta} (z) \right] dz = 0$.  One can further verify that the result of Eq. (\ref{E-field}) is dimensionally correct since $[b] = L ^{-1}$ and $[B] = [E] \, T L ^{-1}$. Finally, since $\mathbf{E} _{i}  \sim \hat{\mathbf{e}} _{z}$ and $\nabla \theta \sim \hat{\mathbf{e}} _{z}$, there are no additional current densities in Eq. (\ref{Modified_Eqs}) and therefore $\mathbf{B} = \mathbf{B} _{0}$.

Once the electromagnetic fields have been determined, we proceed to compute the corresponding electromagnetic potentials in a suitable gauge, so that they can be consistently introduced into the Dirac equation. In the next section we solve analytically the eigenvalue equation related to this fermionic Hamiltonian {in the bulk}, separating electric and magnetic parts. 
Notably, the magnetic sector exhibits supersymmetry (SUSY), whereas the electric sector displays PT-supersymmetry.

\section{Solution to the Dirac-Weyl equation: SUSY and PT-SUSY Hamiltonians}  \label{Dirac_eq_section}

The Dirac-Weyl equation minimally coupled to the electromagnetic field is
\begin{align}
v _{F} \, \boldsymbol{\gamma} \cdot ( - i \hbar \nabla + e \textbf{A}) \, \Psi (\textbf{r},t) - \gamma ^{0} \left( i \hbar \partial _{t} - e  \phi \right) \, \Psi(\textbf{r},t) =0  .
\label{EDIRAC}
\end{align}
where $v _{F}$ is the Fermi velocity and $-e$ is the electron charge. We work in the Weyl (chiral) representation of the Dirac matrices $\gamma^{\mu}$, defined by
\begin{align}
\gamma^{0} = \left( \begin{array}{cc} 0 & \sigma _{0} \\ \sigma _{0} & 0 \end{array} \right), \quad \gamma^{i} = \left( \begin{array}{cc} 0 & \sigma _{i} \\ - \sigma _{i} & 0 \end{array} \right), \quad \gamma ^{5} = \left( \begin{array}{cc} - \sigma _{0} & 0 \\  0 & \sigma _{0} \end{array} \right) . \label{representation}
\end{align}
To proceed further, we have to specify the  potentials associated with the electromagnetic fields computed in the previous section. We work in the Landau gauge so that 
\beq\textbf{A} (y) = A _{x} (y) \hat{\textbf{e}} _{x} = - B _{0} y \, \hat{\textbf{e}} _{x}, \qquad \phi (z) = - \frac{\rho _{0} }{2 \epsilon} z ^{2}.
\label{POTENTIALS}
\eeq
{Since the system remains translationally invariant along the $x$-direction, the corresponding momentum $k_x$ is a good quantum number. Therefore, the dependence on $x$ can be described by plane-wave states. As usual, these states are understood in the sense of box normalization, introducing an auxiliary quantization length that does not affect the final physical results. We therefore introduce the ansatz}
\beq\Psi (\textbf{r},t) = e ^{ i E t / \hbar } \, e ^{ik_{x} x} \, \psi (y,z)
\label{SPINOR0}
\eeq
in Eq. (\ref{EDIRAC}), {obtaining the Dirac equation}
\begin{align}
\hat{\mathcal{H}} \, \psi (y,z) = 0 ,  \label{Weyl_eq3}
\end{align}
{where the Dirac operator is defined by}
\begin{align}
\hat{\mathcal{H}} = \left( \begin{array}{cccc} 0 & 0 & \hat{L} _{\phi} ^{+} &  \hat{L} _{A} ^{+} \\  0 & 0 &   \hat{L} _{A} ^{-} & \hat{L} _{\phi} ^{-} \\  \hat{L} _{\phi} ^{-} & -\hat{L} _{A} ^{+} &  0 & 0 \\ - \hat{L} _{A} ^{-} & \hat{L} _{\phi} ^{+} &  0 & 0  \end{array} \right), \label{Matrix_Hamiltonian}
\end{align}
with
\begin{equation}
\hat{L} _{\phi} ^{\pm}(z)= \mp i \hbar v _{F} \partial _{z} +  E + e  \phi (z),
\qquad 
\hat{L} _{A} ^{\pm} (y) = \mp \hbar v _{F} \partial _{y} + \hbar  v _{F}  k _{x} + e v _{F}  A _{x} (y) . 
\label{Ladder_Operators}
\end{equation}
We now consider  the Dirac  spinor
\begin{align}
\psi (y,z)  = \left( \begin{array}{c} \psi _{l} (y,z)  \\ \psi _{r} (y,z)  \end{array} \right) , 
\end{align}
where
\begin{align}
\psi _{l } (y,z) = \left( \begin{array}{c} \psi _{1} (y,z)  \\ \psi _{2} (y,z)  \end{array} \right),\qquad  \psi _{r} (y,z) = \left( \begin{array}{c} \psi _{3} (y,z)  \\ \psi _{4} (y,z)  \end{array} \right),  \label{left-right_spinors}
\end{align}
are the left-handed and right-handed Weyl components, respectively. They are eigenvectors of the chiral operator $\gamma^5$, with eigenvalues $-1$ and $+1$, respectively.

Before delving into the detailed solutions of the spinors (\ref{left-right_spinors}), let us establish the necessary boundary conditions (BCs) for the problem. Given that the Dirac operator is linear in derivatives, Robin (mixed) BCs are required due to the presence of boundaries in the spatial domain, specifically at $z=\pm L$ in this case. {These boundary conditions define the Dirac problem in the finite slab geometry, ensuring the absence of normal charge flux through the interfaces and determining the corresponding quantized spectrum. It is important to emphasize that the resulting eigenstates remain bulk states extending throughout the interior region $-L<z<L$, rather than surface-localized modes. In particular, the Fermi-arc states associated with the lateral surfaces of the sample are not included in the present analysis.} These BCs are derived in full detail in the Appendix \ref{NORM2} with the results
\begin{eqnarray}
        && \psi_1|_{z=+ L}=-\psi_3|_{z=+ L},\qquad \quad \qquad 
        \psi_2|_{z=+ L}=\psi_4|_{z=+ L}, \label{condition11}\\
        &&\psi_1|_{z=- L}=\psi_3|_{z=- L},\qquad \qquad \qquad
        \psi_2|_{z=- L}=-\psi_4|_{z=- L} ,
        \label{condition21} \\
        && \partial_z\psi_1|_{z=+ L}= \partial_z\psi_3|_{z=+ L},\qquad \qquad  \partial_z\psi_4|_{z=+ L} = \partial_z\psi_1|_{z=- L}, \label{condition31} \\
       &&\partial_z\psi_2|_{z=+ L}=  -\partial_z\psi_3|_{z=- L},\qquad \quad \partial_z\psi_2|_{z=- L}= \partial_z\psi_4|_{z=- L}.
\label{condition41}
\end{eqnarray}

The eigenvalue equation (\ref{Weyl_eq3}) produces the following relations between the components in each sector. For the right sector we have
\beq
\hat{L} _{\phi} ^{+} \psi _{3} (y,z) + \hat{L} _{A} ^{+} \psi _{4} (y,z) = 0 , \qquad \hat{L} _{\phi} ^{-} \psi _{4} (y,z) + \hat{L} _{A} ^{-} \psi _{3} (y,z) = 0 , 
\label{LREL}
\eeq
while for the left sector
\beq
\hat{L} _{\phi} ^{-} \psi _{1} (y,z) - \hat{L} _{A} ^{+} \psi _{2} (y,z) = 0  , \qquad  \hat{L} _{\phi} ^{+} \psi _{2} (y,z) - \hat{L} _{A} ^{-} \psi _{1} (y,z) = 0 . 
\label{RREL}
\eeq
It is important to emphasize that {the components of each}  right and left spinor  {are coupled only through the BCs.} In fact, each pair of equations (\ref{LREL}) and (\ref{RREL}) can be easily decoupled. The same decoupling is obtained by squaring  $\hat{\cal H}$. We obtain 
\begin{equation}
-\hat{\mathcal{H}}^{2}= \left( \begin{array}{cccc} 
\hat{H} _{\phi}^{+}+\hat{H}_{A}^{-}  & 0 & 0 & 0 \\ 0 & \hat{H}_{\phi}^{-} + \hat{H}_{A}^{+} & 0 &  0 \\ 0 & 0 & \hat{H}_{\phi}^{-} + \hat{H}_{A}^{-} & 0  \\ 0 & 0 & 0 & \hat{H}_{\phi}^{+} + \hat{H}_{A}^{+} 
\end{array} \right)=0,
\end{equation}
where we {identify the following contributions, which will be useful when we deal with separation of variables }
\beq
\hat{H} _{A} ^{\pm} = + \hat{L} _{A} ^{\mp} \hat{L} _{A} ^{\pm} =  - \hbar ^{2} v _{F} ^{2} \partial _{y} ^{2} + V_A^{\pm}(y) ,
\qquad 
\hat{H} _{\phi} ^{\pm} = - \hat{L} _{\phi} ^{\pm} \hat{L} _{\phi} ^{\mp} = - \hbar ^{2} v _{F} ^{2} \partial _{z} ^{2} + V_{\phi}^{\pm}(z).
\eeq
In an abuse of notation we call them Hamiltonians.
From  the expressions (\ref{Ladder_Operators}) we identify the corresponding potentials 
\begin{eqnarray}
    V^{\pm}_A(y) =  \left[ \hbar  v _{F}  k _{x} + e v _{F}  A _{x} (y)  \right] ^{2} \pm e \hbar v _{F} ^{2} \partial _{y}  A _{x} (y),
\qquad
    V^{\pm}_\phi(z) =  - \left[ E + e  \phi (z)  \right] ^{2} \pm i e \hbar v _{F} \partial _{z} \phi (z),
\end{eqnarray}
respectively. Notice that the Hamiltonians  $\hat{H} _{A} ^{\pm}$ correspond to Hamiltonians of the type $H^{(1)}$ described in the Appendix  \ref{FACTHAM}. {On the contrary, for the particular electrostatic potential considered in this work, $\hat{H}_{\phi}^{\pm}$ belong to the family $H^{(2)}$ discussed in Appendix A and form a PT-symmetric pair. We emphasize that the original Dirac-Weyl Hamiltonian coupled to real electromagnetic fields remains Hermitian; the PT-symmetric structure arises only at the level of these auxiliary second-order operators introduced through the factorization procedure.} The intertwining operators for the magnetic sector are ${\hat L}_A^{\mp}$, while those of the electric sector are ${\hat L}_\phi^{\mp}$.

\begin{figure}
\begin{center}
        \includegraphics[width=0.9\textwidth]{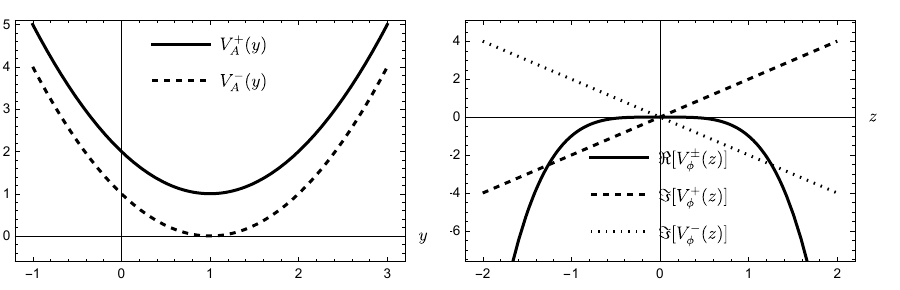}

    \caption{Potentials of the magnetic and electric Hamiltonians for the case $A_x(y) = -B_0y$ and $\phi(z) = -\frac{\rho_0}{2\epsilon}z^2$. The magnetic potentials $V_A^{\pm}(y)$ are purely real and shape invariant (shifted), while the electric potentials $V_\phi^{\pm}(z)$ are the complex  conjugate of each other, exhibiting PT symmetry.}
    \label{figPotentials}
\end{center}    
\end{figure}

Figure \ref{figPotentials} shows the profiles of these potentials for the case to be studied next, with $A_x(y) = -B_0y$ and $\phi(z) = -\frac{\rho_0}{2\epsilon}z^2$. Observe that  that magnetic potentials are purely real $V_A^{\pm}(y)$ and shape invariant (shifted), while the electric potentials  $V_\phi^{\pm}(z)$ are the   complex conjugate of each other, exhibiting PT symmetry.
Now, since each operator $\hat{H} _{\phi} ^{\pm}$ commutes with any other $\hat{H} _{A} ^{\pm}$, 
 one can use separation of variables to satisfy the eigenspinor equation $\hat{\mathcal{H}} ^{2} \psi (y,z) = 0 $. As a first step to  implement separation of variables we define  
\beq
\psi _{l} (y,z) = \left( \begin{array}{c} \mathcal{A}\mathcal{Z}_l^{+} (z) \,  \mathcal{Y} ^{-} (y) \\ \mathcal{B} \mathcal{Z}_l ^{-} (z) \,  \mathcal{Y} ^{+} (y) \end{array} \right), \qquad 
\psi _{r} (y,z) = \left( \begin{array}{c} \mathcal{C} \mathcal{Z}_r ^{-} (z) \,  \mathcal{Y} ^{-} (y) \\ \mathcal{D} \mathcal{Z}_r ^{+} (z) \,  \mathcal{Y} ^{+} (y) \end{array} \right) ,
\label{SPINORS}
\eeq
where $\mathcal{A},\mathcal{B},\mathcal{C}$ and $\mathcal{D}$ are unknown amplitudes of the total spinor $\psi(y,z)$  to be determined later. We find necessary to  consider  two independent functions $Z_l$ and  $Z_r$ for each Hamiltonian $H_\phi$, while we do not distinguish $l$ and $r$ in the $y$ coordinate.

The second step requires to introduce   the eigenvalue equations
\beq
\hat{H} _{\phi} ^{\pm} \mathcal{Z}_{l,r} ^{\pm} (z) = E _{\phi} ^{\pm} \, \mathcal{Z}_{l,r} ^{\pm} (z) , \qquad  \hat{H} _{A} ^{\pm} \mathcal{Y} ^{\pm} (y) = E _{A} ^{\pm} \, \mathcal{Y} ^{\pm} (y).
\label{EIGENEQS}
\eeq
As shown in the Appendix \ref{FACTHAM}, each pair of Hamiltonians $\hat{H} _{\phi} ^{\pm} $ and $\hat{H} _{A} ^{\pm}$ is isoespectral, such that $E _{\phi} ^{\pm} \equiv E _{\phi}$
and $E _{A} ^{\pm} \equiv E _{A}$, except for a possibly unpaired  ground state in each case.
Applying the operator ${\cal H}^2$ to the spinors $\Psi_L$ and $\Psi_R$ in (\ref{SPINORS}) we get the condition
\begin{align}
 E _{\phi}  + E _{A} = 0, 
 \label{CONDE}
\end{align}
in every entry. This is an important constraint which demands to find   solutions for the PT-symmetric Hamiltonians  $\hat{H} _{\phi} ^{\pm}$ {that have real energies}, which is not guaranteed beforehand.

\subsection{The magnetic sector: Conventional SUSY QM}

Now we solve the eigenvalue equation for the magnetic sector. First, we observe that introducing the superpotential
\begin{align}
W_A (y) = \hbar  v _{F}  k _{x} + e v _{F}  A _{x} (y)  , 
\label{SUPPOTA}
\end{align}
the corresponding Hamiltonian can be written as
\begin{align}
\hat{H} _{A} ^{\pm}=+ \hat{L} _{A} ^{\mp} \hat{L} _{A} ^{\pm}  &= - \frac{\partial ^{2}}{\partial \tilde{y} ^{2}}  + W_A ^{2} (y) \pm \frac{\partial W_A  (y) }{\partial \tilde{y}},
\label{FACTHA}
\end{align}
where $\tilde{y} = y / \hbar v _{F}  $.  As discussed in the Appendix \ref{FACTHAM}, we recognize the Hamiltonians $\hat{H} _{A} ^{+}$ and $\hat{H} _{A} ^{-}$ as supersymmetric partners.  The corresponding eigenfuntions ${\cal Y}^{\pm}$ were defined in Eq. (\ref{EIGENEQS}), 
As shown in the Appendix \ref{FACTHAM}, it is simply the factorization of both $H_A^\pm$, as  indicated in  Eq. (\ref{FACTHA}), which yields the  intertwining relations
 \beq
 {\hat L}^+_A \, {\cal Y}_A^+= v_+ \sqrt{E_A}\,  {\cal Y}_A^-, \qquad
{\hat L}^-_A \, {\cal Y}_A^-= v_- \sqrt{E_A} \,{\cal Y}_A^+,\qquad v_+ v_-=1.
 \label{INTERTWA}  
 \eeq
We now take $A _{x} (y) = - B _{0} y $ in Eq. (\ref{SUPPOTA}) and solve the eigenvalue equation
\begin{align}
\left[ - \hbar ^{2} v _{F} ^{2} \frac{\partial ^{2}}{\partial y ^{2}}  + ( \hbar  v _{F}  k _{x} - e v _{F} B _{0} y) ^{2} \mp e v _{F}  ^{2} \hbar B _{0} \right] \mathcal{Y} ^{\pm} (y) = E _{A} ^{\pm} \, \mathcal{Y} ^{\pm} (y)  .
\end{align}
After simple manipulations we cast this equation into the quantum harmonic oscillator form:
\begin{align}
\left[ - \frac{\hbar ^{2}}{2m}  \frac{\partial ^{2}}{\partial y ^{2}}  \right. + \left. \frac{1}{2} m \omega_B ^{2}  \left( y - y _{0} \right) ^{2} \right] \mathcal{Y} ^{\pm} (y) = \left[ E _{A} ^{\pm} \pm \frac{\hbar\omega_B}{2} \right] \, \mathcal{Y} ^{\pm} (y)  ,
\end{align}
where $y _{0} \equiv \frac{ \hbar   k _{x} }{ e  B _{0} } $, $2m = 1 / v _{F} ^{2} $ and $\omega_B = 2 e v _{F} ^{2} B _{0}$. Therefore, the normalized energy eigenfunctions read
\begin{align}
\mathcal{Y} ^{\pm} _{n _{\pm}} (y) = \frac{1}{\sqrt{2^{n_{\pm}} n_{\pm} ! }} \left( \frac{1}{\pi l_{B}^{2}} \right)^{\frac{1}{4}} e^{-\frac{1}{2} \left( \frac{y - y_{0}}{l_{B}} \right) ^{2} } H_{n_{\pm}} \left( \frac{y- y_{0}}{l_{B}} \right),
\end{align}
where $H_n$ are the Hermite polynomials, $l _{B} ^{2} = \hbar / ( e  B _{0} )$ is the magnetic length and the corresponding energy levels are
\begin{align}
E _{A \, n _{\pm}} ^{\pm}  = \hbar \omega_B \left(n _{\pm} + \frac{1}{2} \mp \frac{1}{2} \right) , \qquad n _{\pm} = 0,1,2, \cdots . 
\end{align}
Since isospectrality yields  $E _{A \, n _{+}} ^{+} = E _{A \, n _{-}} ^{-} = E_{A\, n}$, we conclude that $n _{+} = n _{-} + 1$. In the following, we take $n _{+} \equiv n$, such that $\{ E _{A \, n} ^{+} \}$, with $n =0,1,2, \cdots $, is the discrete spectrum of $\hat{H} _{A} ^{+}$. The ground state of $\hat{H} _{A} ^{+}$ is annihilated by $\hat{L} _{A} ^{+}$: 
\begin{align}
\hat{L} _{A} ^{+} \, \mathcal{Y} ^{+} _{0} (y) = 0  , 
\end{align}
and hence the ground state eigenvalue of $\hat{H} _{A} ^{+}$ is $E _{A \, 0} ^{+} = 0$. Therefore, the discrete spectrum of $\hat{H} _{A} ^{-}$ consists of the eigenvalues $\{ E _{A \, n} ^{-}  \}$ and normalized eigenfunctions $\{ \mathcal{Y} ^{-} _{n} (y) \}$. Following the standard phase convention in the harmonic oscilator we take $v_+=v_-=1$ and we have
\begin{align}
\label{IntertMagn}
\mathcal{Y}^{-} _{n-1} (y) = \frac{1}{\sqrt{ E _{A \, n} }} {\hat L} _{A} ^{+} \, \mathcal{Y} ^{+} _{n} (y), \qquad 
\mathcal{Y}^{+} _{n} (y) = \frac{1}{\sqrt{ E _{A \, n} }} {\hat L} _{A} ^{-} \, \mathcal{Y} ^{-} _{n-1} (y)  , \qquad n =1,2,3, \cdots . 
\end{align}

\subsection{The electric sector: a PT-symmetric Hamiltonian.}

Now we solve the eigenvalue equation for the electric sector. Proceeding in analogy to the supersymmetric case we introduce the complex superpotential
\begin{align}
W_{\phi}(z) = i \left[ E + e  \phi (z)  \right]
\end{align}
such that
\begin{align}
\hat{H} _{\phi} ^{\pm}= - \hat{L} _{\phi} ^{\pm} \hat{L} _{\phi} ^{\mp} = - \hbar ^{2} v _{F} ^{2} \frac{\partial ^{2}}{\partial z ^{2}} + W_{\phi} ^{2} (z)  \pm   \hbar v _{F} \frac{\partial W_{\phi} (z)  }{\partial z } .
\end{align}
From the imaginary superpotential we recognize the Hamiltonians $\hat{H} _{\phi} ^{+}$ and $\hat{H} _{\phi} ^{-}$ as PT-supersymmetric partners. Again, the factorization of $H_\phi^\pm $ produces the intertwining relations
\beq
{\hat L}^-_\phi {\cal Z}^+= s_- \sqrt{E_A} {\cal Z}^-, \qquad
{\hat L}^+_\phi {\cal Z}^-= s_+ \sqrt{E_A} {\cal Z}^+, \qquad s_+ s_-=1.
\label{INTERTWB}
\eeq
where we have used the relation $E_\phi=-E_A$.  The values of $s_+$ and $s_-$ are determined in the Appendix \ref{INTERT} and we obtain
{\beq
s_+=i \kappa  , \qquad s_-=-i \kappa, \qquad \kappa= \pm 1,
\eeq
such that we have two choices for the pair $(s_+, s_-)$. }
We now take $\phi (z) = - \frac{\rho _{0} }{2 \epsilon} z ^{2} $ and solve the eigenvalue equation 
\begin{align}
\left[ -  \frac{\partial ^{2}}{\partial \tilde{z} ^{2}} - \left( \epsilon - \lambda \tilde{z} ^{2}  \right) ^{2} \mp  2 i \lambda \tilde{z} \right] \mathcal{Z} ^{\pm} (z) = ( E _{\phi} ^{\pm} / \mathcal{E} ^{2} ) \, \mathcal{Z} ^{\pm} (z) . 
\label{EQSCHZ}
\end{align}
where $\tilde{z} \equiv z/L \in [-1,1]$, $\varepsilon \equiv E / \mathcal{E}$ and $\mathcal{E} \equiv \hbar v _{F} / L$ (with dimensions of energy). Also, we introduce the dimensionless parameter
\begin{align}
\lambda = \frac{L ^{3} b}{l _{B} ^{2} } \frac{  \alpha}{\pi \epsilon _{r} \beta } , 
\end{align}
where $\alpha = \frac{1}{4 \pi \epsilon _{0}} \frac{e ^{2}}{\hbar c} \approx 1/137 $ is the fine structure constant, $\epsilon _{r} = \epsilon / \epsilon _{0} > 1$ is the relative permittivity and $\beta \equiv v _{F}/c < 1$. Using the fact that $E _{\phi} ^{\pm} = E _{\phi} = - E _{A} = - 2 e \hbar v _{F} ^{2} B _{0} n$, we get
\begin{align}
\left[ -  \frac{\partial ^{2}}{\partial \tilde{z} ^{2}} - \left( \varepsilon -   \lambda \tilde{z} ^{2}  \right) ^{2} \mp 2i \lambda \tilde{z} + 2 n \ell ^{2} \right] \mathcal{Z} ^{\pm} (\tilde{z}) = 0 , \label{Z_diff_eq}
\end{align}
where $\ell = L / l _{B}$. {An important feature of Eq.~(\ref{Z_diff_eq}) is that the dimensionless energy $\varepsilon$ appears both as a spectral parameter and within the effective potential. Consequently, the equation is energy-dependent and the admissible values of $\varepsilon$ must be determined together with the corresponding boundary conditions. In general, energy-dependent eigenvalue problems may require modified normalization prescriptions and inner products~\cite{Formanek2004}. In the present case, however, Eq.~(\ref{Z_diff_eq}) arises as an auxiliary equation obtained from the factorization of the original Dirac-Weyl problem. The physical states are the complete Dirac spinors, and their normalization is ultimately imposed on the full spinor solution, as discussed below in Eq.~(60).

}

The differential equation (\ref{Z_diff_eq}), is a particular case of $y'' + (\gamma + \delta z + \epsilon z ^{2}) y' + (\alpha z - q) y  = 0$, which is solved in terms of the tri-confluent Heun function $\mbox{HeunT} [q,\alpha , \gamma , \delta , \epsilon , z]$ \cite{ronveaux1995heun}. The explicit linearly independent solutions read
\begin{align}
u _{1} ^{\pm} (\tilde{z}) &= e ^{ \pm i \tilde{z} \left( \varepsilon - \frac{1}{3} \lambda \tilde{z} ^{2} \right) } \,  \mbox{HeunT} [2 n \ell ^{2} , 0 , \pm 2 i \varepsilon , 0 , \mp 2 i \lambda , \tilde{z} ] , \label{U1}\\[5pt] u _{2} ^{\pm} (\tilde{z}) &= e ^{ \mp i \tilde{z} \left( \varepsilon - \frac{1}{3} \lambda \tilde{z} ^{2} \right) } \,  \mbox{HeunT} [2 n \ell ^{2} , \pm 4 i \lambda , \mp 2 i \varepsilon , 0 , \pm 2 i \lambda , \tilde{z} ]  \label{U2}.
\end{align}
One can further verify that these functions satisfy the symmetry properties 
\beq
u _{1} ^{\pm} (\tilde{z}) = u _{1} ^{\mp} (- \tilde{z}) = [u _{1} ^{\pm} (-\tilde{z})] ^{\ast},
\qquad
u _{2} ^{\pm} (\tilde{z}) = u _{2} ^{\mp} (- \tilde{z}) = [u _{2} ^{\pm} (- \tilde{z} )] ^{\ast}, \label{u_properties}
\eeq
which could  be inferred from the differential equation  (\ref{Z_diff_eq}), but only up to a constant factor.   {
 Motivated by the relations  (\ref{u_properties}) we also demand 
\begin{align}
\mathcal{Z} ^{\pm} (\tilde{z}) = \mathcal{Z} ^{\mp} (- \tilde{z}) , \qquad [ \mathcal{Z} ^{\pm} (\tilde{z}) ] ^{\ast } = \mathcal{Z} ^{\pm} (-\tilde{z})  . \label{Z_properties}
\end{align}
As shown in the Appendix \ref{APPD} the relations (\ref{Z_properties}) are fulfilled provided the coefficients of the expansion in terms of $u_1$ and $u_2$ are real. {In the Appendices \ref{APPC11}  and   \ref{APPC2}    we show how the  BCs upon the spinor components (\ref{condition11})-(\ref{condition41}}) translate into conditions over the functions ${\cal Z}^{\pm}_{l,r}$ at the interfaces. The symmetries (\ref{Z_properties}) allow reducing all BCs to those at the interface ${\tilde z}= + 1$, which are finally summarized in
\beq
\Big({\cal Z}_l^+ + i\eta \, [{\cal Z}_r^+]^*\Big)_{{\tilde z}=+1}=0, \qquad  \Big(\partial_z{\cal Z}_l^+ - i\eta \,  [\partial_z{\cal Z}_r^+]^*\Big)_{{\tilde z}=+1}=0, \qquad \eta=\pm 1.
\label{FINALBC12}
\eeq
Also, the independent constants in the spinors (\ref{SPINORS}) become related as follows
\beq
{\cal C}=i \eta {\cal A}, \qquad  {\cal D}=i \eta {\cal B}, 
\label{CD}
\eeq
which defines  the  quantum number $\eta$,  leaving  us with two arbitrary constants ${\cal A}$ and  ${\cal B}$. {A further reduction is obtained by making sure   that  the ansatz  (\ref{SPINORS}) indeed satisfies the Dirac equation (\ref{EDIRAC}). This calculation is presented in the Appendix \ref{APPE}. It should be noted that  explicit expressions for the functions ${\cal Y^{\pm}}$ and ${\cal Z}^{\pm}_{l,r}$ are not required. Only a judicious rewriting  of the derivatives $\partial_y$  and $\partial_z$ in terms of the operators ${\hat L}^{\pm}_A$ and ${\hat L}^{\pm}_\phi$, plus extensive use of the intertwining relations (\ref{INTERTWA}) and (\ref{INTERTWB}) is necessary.  We obtain the additional  relations}
{\beq
{\cal A}= i \kappa  {\cal B}, \qquad  {\cal C}= i \kappa  {\cal D}, \qquad \kappa= \pm 1, 
\label{AC}
\eeq
introducing another label $\kappa$. 
Using  Eqs. (\ref{CD}) and  (\ref{AC}) we express all coefficients in terms of ${\cal A}$ yielding 
\beq
{\cal B}= -i \kappa {\cal A}, \qquad {\cal C}=i\eta {\cal A}, \qquad {\cal D}=  \kappa \eta {\cal A}.
\eeq
The remaining constant ${\cal A}$ is determined by normalization of the full spinor $\Psi$ using the standard Dirac probability density. 
At this stage the  solution of the Dirac equation is    
\begin{equation}
   \Psi=  \begin{pmatrix}
        \psi_1\\
        \psi_2\\
        \psi_3\\
        \psi_4
    \end{pmatrix} = \mathcal{A} \, e ^{i E t/\hbar}  e^{ik_x x}\begin{pmatrix}
        \mathcal{Z}_{l}^+({\tilde z}) \mathcal{Y}^-(y)\\
        -i \kappa\mathcal{Z}_l^-({\tilde z}) \mathcal{Y}^+(y)\\
        i \eta \,\mathcal{Z}_r^-({\tilde z}) \mathcal{Y}^-(y)\\
         \eta \kappa \, \mathcal{Z}_r^+({\tilde z}) \mathcal{Y}^+(y)
    \end{pmatrix}.
    \label{GENSPINOR21}
\end{equation}
Some comments regarding the physical identification of the additional quantum numbers $\eta$ and $\kappa$ are provided in  section \ref{DEG} of the manuscript.}

The next step is to construct the specific solutions for the $z$-dependent contributions  in the spinors (\ref{SPINORS}) writing
\beq
{\cal Z}^+_l({\tilde z})=a_{1l}\, u^+_1({\tilde z})+ a_{2l}\, u^+_2({\tilde z}), \qquad 
{\cal Z}^+_r({\tilde z})=a_{1r}\, u^+_1({\tilde z})+ a_{2r}\, u^+_2({\tilde z}),
\label{DEFZRZL1}
\eeq
which include the four real coefficients $a_{k l}, a_{k r}$, where $k=1,2$,  to be determined. The corresponding functions  ${\cal Z}^-(z)$ can be obtained from the relations (\ref{Z_properties}). We substitute the ansatz (\ref{DEFZRZL1}) into the BCs (\ref{FINALBC12}) and separate real and imaginary parts in each equation yielding four conditions upon the coefficients, which are  presented as 
\begin{equation}
\begin{pmatrix}
R_1 & \eta I_1 & R_2 & \eta I_2\\
I_1 & \eta R_1 & I_2 & \eta R_2\\
R'_1 & -\eta I'_1 & R'_2& -\eta I'_2\\
I'_1 & -\eta  R'_1 & I'_2& -\eta R'_2
\end{pmatrix}
\begin{pmatrix}
        a_{1l}\\
        a_{1r}\\
        a_{2l}\\
        a_{2r}
    \end{pmatrix} =0,
    \label{DISPREL0}
\end{equation}
with the notation
\beq
u_k({\tilde z}=1)=R_k+i I_k, \qquad [\partial_z u_k(z)]_{{\tilde z}=+1}= R'_k+i I'_k, \qquad k=1,2.
\eeq
in terms of the real and imaginary parts of the functions
$u_k$ and $\partial_z u_k$ evaluated at ${\tilde z}=+1$.
{As can be directly seen from the second columm of the  expressions (\ref{condition11})-(\ref{condition41}) for the BC's, the sector  $\psi_2$ is linearly related to $\psi_4$. Since both components are proportional to $\kappa$, the BC's  turn out to be  independent of this label.}

Calling $M$ the matrix in Eq. (\ref{DISPREL0}), we require
$\det M=0$, which defines the  dispersion relation of  the system yielding the eigenvalues for the energy $E$. As shown in the Appendix \ref{APPD}  this condition requires to find the zero's of the factors $Z1$ and $Z2$ defined in Eq. (\ref{Z1Z2}). An important property is that $\det M$ is independent of $\eta=\pm 1$. This introduces a two-fold degeneracy in the spectrum since the remaining coefficients, defining the associated wave function, will be $\eta$-dependent. In the rest of the paper  we consider only the  choice $\eta=+1$.

From now on, the calculation is carried out numerically. Next, for a given energy eigenvalue, we determine the coefficients $a_{k l}, a_{k r}$ by verifying that the matrix has a rank of $3$ in each case, and solving for three of the coefficients in terms of one of them, say 
$a_{1l}$, for example, which remains the only undetermined constant related to the ${\cal Z}^{\pm}_{l,r}$ functions. 
Since we still have the overall coefficient ${\cal A}$ in the spinor, we take $a_{1l}=1 $   with no loss of generality, as this amounts to redefine   ${\cal A} \, a_{1l} \to {\cal A} $,   and finally impose normalization on the only remaining arbitrary coefficient. In other words, we still need to fix the constant $\mathcal{A}$, such that 
\begin{equation}
\int^{+\infty}_{-\infty} dy \int^{+1}_{-1{}} d{\tilde z} \, \, \Psi^\dagger\,  \Psi=1 
\label{NORMALIZATION}
\end{equation}
with $\Psi$ given by Eq. (\ref{GENSPINOR21}). Nevertheless, as there is not analytic expression for the normalization of the Heun's functions, Eq. (\ref{NORMALIZATION}) is numerically evaluated.

\begin{figure}
    \centering
    \includegraphics[width=0.49\linewidth]{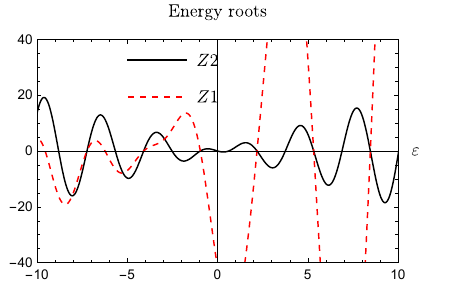}
    \caption{The dimensionless energy eigenvalues $\varepsilon_{n,m}$, for $n=0$, for the conditions $Z2=0$ (solid, black lines) and $Z1=0$ (dashed, red lines). The parameters are   $B_0=1.0\,$T, $b=0.26\,$nm$^{-1}$ and $L=100\,$nm.}
    \label{energies}
\end{figure}

\begin{figure}
    \centering
    \includegraphics[width=0.98\linewidth]{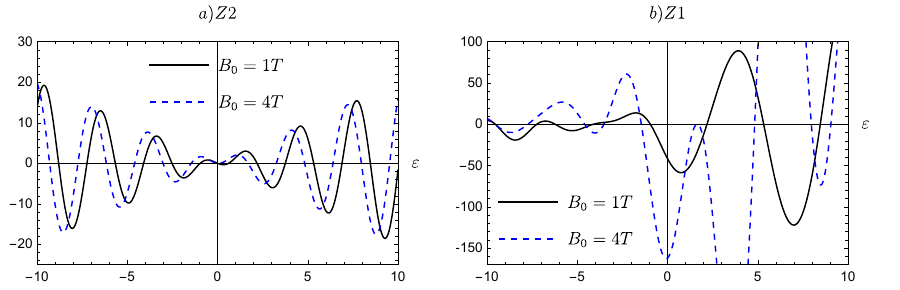}
    \caption{The dimensionless energy eigenvalues $\varepsilon_{n,m}$, for $n=0$, for the conditions $a$) $Z2=0$ and $b$) $Z1=0$, for magnetic fields $B_0=1$T (solid, black lines) and $B_0=4$T (dashed, blue lines), respectively. The remaining parameters are $b=0.26\,$nm$^{-1}$ and $L=100\,$nm.}
    \label{energiesB}
\end{figure}

{As previously stated, the condition $\det M= Z1 \times Z2 =0$ determines the energy spectrum of the system, $\{E_{n,m}\}$ which we write in units  of the energy  ${\cal E}=\hbar v_F/L $ as  $E_{n,m}= \varepsilon_{n,m} \, {\cal E}$.}} We have taken the values $v_F= 5 \times 10 ^{5}$m/s, $b=0.26\,$nm$^{-1}$ and $L=100\,$nm. For fixed $n$ the zero determinant  condition produces an infinite number of root energies $\varepsilon_{n,m}$, \textit{i.e.} intersections of the functions $Z1$ and $Z2$ with the $\varepsilon$-axis, which define the root quantum number $m=0,\pm 1,\pm 2,...$ 
In our tables and figures we present the spectrum as the set of values {$ \{ \varepsilon_{n,m} \}$.

Figure \ref{energies} shows the dimensionless {energies $\varepsilon_{n,m}$} for $B_0=1.0\,$T, determined by the points where the functions {$Z2$ and $Z1$ (solid and dashed lines, respectively) intersect the $\varepsilon$-axis for $n=0$, where $n$ is the principal quantum number. Figure \ref{energiesB} shows the dimensionless {energies $\varepsilon_{n,m}$}, determined by the points where the functions {$Z2$ [Fig. \ref{energiesB}$a$)] and $Z1$ [Fig. \ref{energiesB}$b$)]  intersect the $\varepsilon$-axis for  $n=0$. Each figure includes the choices $B_0=1.0\,$T (solid line) and $B_0=4.0\,$T (dashed line) for the magnetic field.}

In Tables \ref{tab:Evsn_Z2} and \ref{tab:Evsn_Z1} we list the dimensionless energies {$\varepsilon_{n,m}$,} for $m=0, \pm 1,  \pm 2$, for some values of $n$ and $B_0=1$T,  arising from $Z2=0$ and $Z1=0$ (no $m=0$ level in this case), rspectively. The same numerical values of the parameters in Fig. \ref{energies} were used.

In Fig. \ref{Evsn}$a$)  we show the complete spectrum (even and odd $n$) for different root numbers, $m$, for both $Z2=0$ (solid lines) and $Z1=0$ (dashed lines) conditions. From this figure it can be noted an irregular behavior of the roots with $n$ for both cases. Also, some roots of $Z1=0$ coincide with roots of $Z2=0$. 
On the other hand, from Figs. \ref{Evsn}$b$), the even and odd spectrum of $Z2=0$ (solid and dashed lines, respectively) present an approximate linear behavior specially as $m$ grows. 

 {In addition, Tables \ref{tab:EvsB_Z2} and \ref{tab:EvsB_Z1} present the dimensionless root energies {$\varepsilon_{n,m}$,} $m=0, \pm 1,  \pm 2$, for $n=0$,  arising from $Z2=0$ and $Z1=0$ (no $m=0$ level in this case), respectively, varying the magnetic field values from $1.0$-$5.0$T.

{Figure \ref{EvsB}$a$) shows the dimensionless root energy $\varepsilon_{n,m}$ as a function of the root quantum number $m$ for different magnetic field intensities $B_0$ setting $n=0$, coming from conditions $Z2 = 0$ (solid, black lines) and $Z1 = 0$ (dashed, redlines). We observe an increasing linear behavior of the root energy with $m$, such that the energy decreases as the magnetic field increases. In that sense, for sufficiently large magnetic fields the lowest nonzero energy states can become negative, as shown in the inset figure, for the case of $B_0=5.0$T and $m=1$. Also, as previously discussed, some values of $Z1 = 0$ coincide with some of $Z2 = 0$. Figure \ref{EvsB}$b$) shows with more clarity the transition from positive to negative energies as $B_0$ grows, where energy values are shown as a function of $B_0$ for $m=-2,-1,1,2$.}

\begin{table}[htbp]
    \centering
    \begin{tabular}{|c|c|c|c|c|c|}
        \hline
        $n$ & $\varepsilon_{n,-2}$ & $\varepsilon_{n,-1}$ & $\varepsilon_{n,0}$ & $\varepsilon_{n,1}$ & $\varepsilon_{n,2}$ \\
        \hline
        $0$ & $-2.5227$ & $-0.9519$ & $0$ & $0.6188$ & $2.1896$ \\
        $1$ & $-3.4400$ & $-1.6803$ & $0$ & $0.5685$ & $3.3963$ \\
        $2$ & $-3.1860$ & $-0.9564$ & $0$ & $1.7976$ & $6.1541$ \\
        \hline
    \end{tabular}
    \caption{The dimensionless energy eigenvalues $\varepsilon_{n,m}$, for $m=0, \pm 1,  \pm 2$, for some values of $n$,  arising from $Z2=0$. The parameters are  $B_0=1.0\,$T, $b=0.26\,$nm$^{-1}$ and $L=100\,$nm.}
    \label{tab:Evsn_Z2}
\end{table}


\begin{table}[htbp]
    \centering
    \begin{tabular}{|c|c|c|c|c|}
    \hline
       $n$ &  $\varepsilon_{n,-2}$ &  $\varepsilon_{n,-1}$ &  $\varepsilon_{n,1}$ & $\varepsilon_{n,2}$ \\
      \hline   
        $0$   &  $-4.0935$ &  $-0.9519$  &  $2.1896$ &  $5.3312$ \\    
        $1$  &  $-3.0876$ &  $-1.0753$  &  $0.5463$ &  $5.9942$ \\    
        $2$  &  $-2.9289$ &  $-0.8796$  &  $3.7296$ &  $10.7440$ \\   
        $3$  &  $-4.3491$ &  $-2.1030$  &  $1.7737$ &  $4.6905$ \\
        $4$   &  $-5.5666$ &  $-3.1987$  &  $0.2708$ &  $3.3258$ \\
       \hline
    \end{tabular}
    \caption{The dimensionless energy eigenvalues $\varepsilon_{n,m}$, $\varepsilon_{n,m}$, for $m= \pm 1,  \pm 2$, for some values of $n$,  arising from $Z1=0$. The parameters are  $B_0=1.0\,$T, $b=0.26\,$nm$^{-1}$ and $L=100\,$nm.}
    \label{tab:Evsn_Z1}
\end{table}

\begin{figure*}[!tb]
  \begin{center}
            \includegraphics[width=1\textwidth]{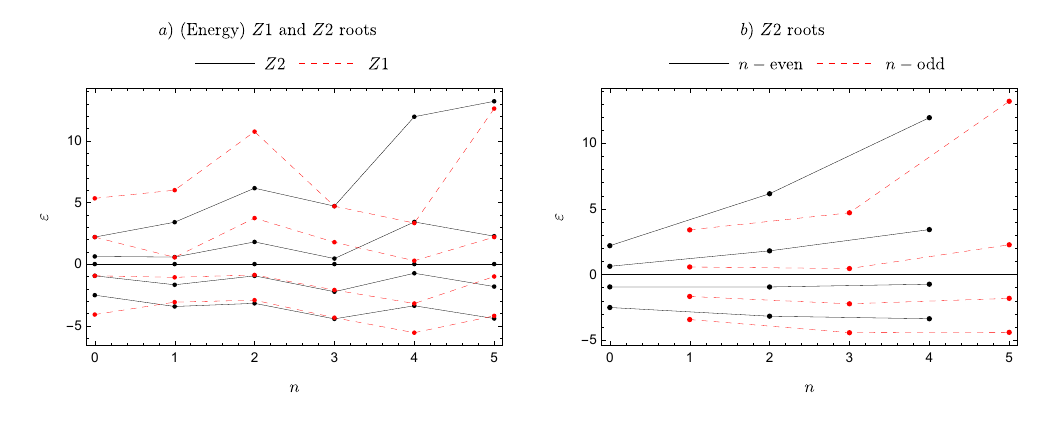}
        
    \caption{{The dimensionless energy eigenvalues $\varepsilon_{n,m}$, for $m=-2,-1,0,1,2$ (in ascendant order) as a function of the  quantum number $n$. $a$) Complete spectrum (even and odd $n$). Black solid lines represent the roots of the condition $Z2=0$, while red dashed lines represent the roots of the condition $Z2=0$. $b$) Even $n$ (black solid) and odd $n$ (red dashed) coming from the condition $Z2=0$. In this plot the $m=0$ roots are not included to avoid image saturation.}}
    \label{Evsn}
  \end{center}
\end{figure*}

\begin{table}[htbp]
    \centering
    \begin{tabular}{|c|c|c|c|c|c|}
    \hline
       $B_0$(T) &  $\varepsilon_{n,-2}$ &  $\varepsilon_{n,-1}$ &  $\varepsilon_{n,0}$ &  $\varepsilon_{n,1}$ & $\varepsilon_{n,2}$ \\
      \hline   
        $1.0$  &  $-2.5227$ &  $-0.9519$ &  $0$ &   $0.6188$ &  $2.1896$ \\    
        $2.0$  &  $-2.6315$ &  $-1.1184$ &  $0$ &  $0.4523$ &  $2.0231$ \\    
        $3.0$  &  $-2.8557$ &  $-1.2849$ &  $0$ &  $0.2858$ &  $1.8566$\\   
        $4.0$  &  $-3.0222$ &  $-1.4514$ &  $0$ &  $0.1193$ &  $1.6901$ \\
        $5.0$  &  $-3.1887$ &  $-1.6179$ &  $0$ &  $-0.0471$ &  $1.5236$ \\
       \hline
    \end{tabular}
    \caption{The dimensionless energy eigenvalues ${\varepsilon_{n,m}}$, for $m=0,\pm1,\pm2$, for some values of $B_0$ with $n=0$, arising from $Z2=0$. The parameters used are $b=0.26\,$nm$^{-1}$ and  $L=100\,$nm.}
    \label{tab:EvsB_Z2}
\end{table}

\begin{table}[htbp]
    \centering
    \begin{tabular}{|c|c|c|c|c|c|}
    \hline
       $B_0$(T) &  $\varepsilon_{n,-2}$ &  $\varepsilon_{n,-1}$  &  $\varepsilon_{n,1}$ & $\varepsilon_{n,2}$ \\
      \hline   
        $1.0$  &  $-4.9035$ &  $-0.9519$  &   $2.1896$ &  $5.3312$ \\    
        $2.0$  &  $-4.2600$ &  $-1.1184$  &  $2.0231$ &  $5.1647$ \\    
        $3.0$  &  $-4.4265$ &  $-1.2849$  &  $1.8566$ &  $4.9982$\\   
        $4.0$  &  $-4.5930$ &  $-1.4514$  &  $1.6901$ &  $4.8317$ \\
        $5.0$  &  $-4.7595$ &  $-1.6179$  &  $1.5236$ &  $4.6651$ \\
       \hline
    \end{tabular}
    \caption{The dimensionless energy eigenvalues ${\varepsilon_{n,m}}$, for $m=\pm1,\pm2$, for some values of $B_0$ with $n=0$, arising from $Z1=0$. The parameters used are $b=0.26\,$nm$^{-1}$ and  $L=100\,$nm.}
    \label{tab:EvsB_Z1}
\end{table}

\begin{figure}
\begin{center}
        \includegraphics[width=01.\textwidth]{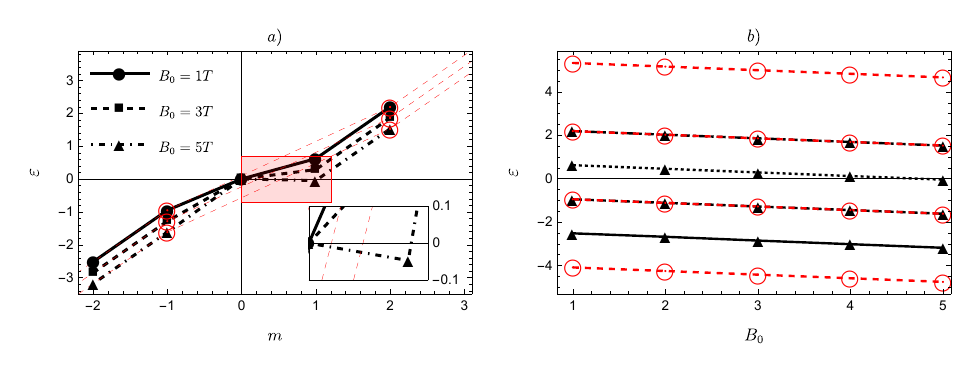}
    
    \caption{{$a$) The dimensionless energy eigenvalues $\varepsilon_{n,m}$ as a function of the root quantum number $m$, for different values of the magnetic field $B_0=1$T, $3$T, $5$T, coming from conditions $Z2=0$ (black) and $Z1=0$ (red). Inset: Magnification of the red rectangle area, showing a negative root-energy value for $m=1$ for $B_0=5$T. $b$) $\varepsilon_{n,m}$ as a function of the magnetic field $B_0$ for $m=-3,-2,-1,1,2,3$, coming from conditions $Z2=0$ (black) and $Z1=0$ (red). In both figures we have set $n=0$.}}
    \label{EvsB}
\end{center}    
\end{figure}

In the next section, we compute the corresponding probability densities and current densities of the system in order to study  its electronic and transport properties. 

\section{Dynamics of the Weyl fermions} \label{currents_section}

\subsection{ {Probability densities and currents} }

\begin{figure*}[!tb]
  \begin{center}
        \includegraphics[width=0.98\textwidth]{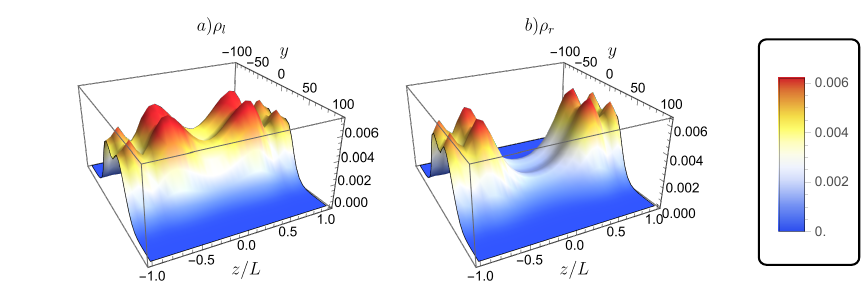}
    \caption{Probability density {for each chirality. $a$) Left chirality and $b$) right chirality.}}
    \label{probability}
  \end{center}
\end{figure*}

\begin{figure*}[!tb]
  \begin{center}
        \includegraphics[width=0.98\textwidth]{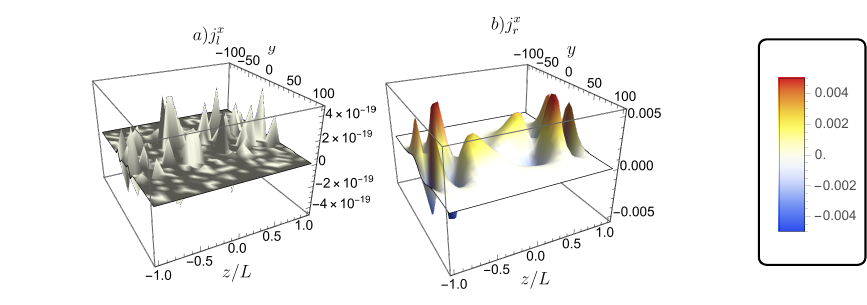}
        
    \caption{The $x$-component of the probability density current {for each chirality. $a$) Left chirality and $b$) right chirality.}}
    \label{currentX}
  \end{center}
\end{figure*}

\begin{figure*}[!tb]
  \begin{center}
    
        \includegraphics[width=0.98\textwidth]{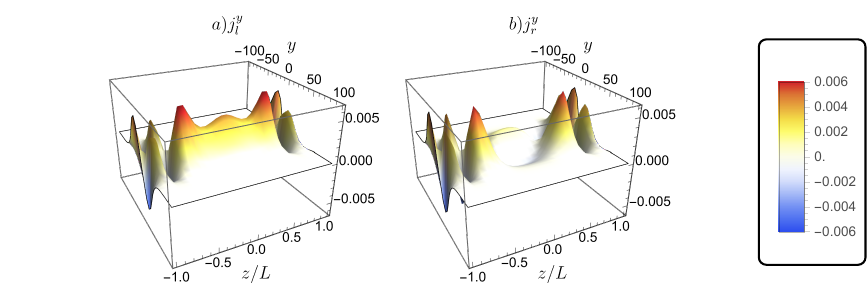}
        
    \caption{The $y$-component of the probability density current {for each chirality. $a$) Left chirality and $b$) right chirality.}}
    \label{currentY}
  \end{center}
\end{figure*}

\begin{figure*}[!tb]
  \begin{center}
    
        \includegraphics[width=0.98\textwidth]{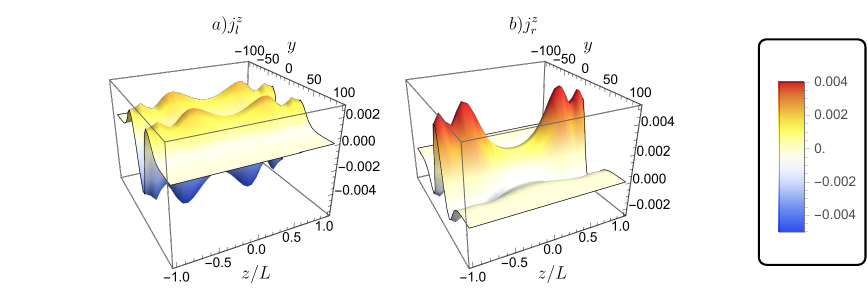}
        
    \caption{The $z$-component of the probability density current {for each chirality. $a$) Left chirality and $b$) right chirality.}}
    \label{currentZ}
  \end{center}
\end{figure*}

\begin{figure*}[!tb]
  \begin{center}
    
        \includegraphics[width=0.9\textwidth]{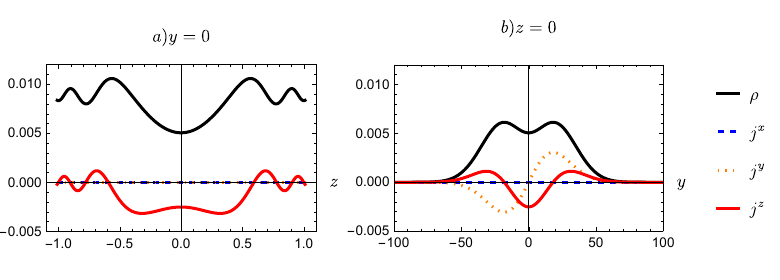}
        
    \caption{Probability density and currents for the values a) $y=0$ and b) $z=0$.}
    \label{slices}
  \end{center}
\end{figure*}

In order to gain insight of the electronic and transport properties of the system, we compute the probability density $\rho(\mbf{x}, t)$ and the  probability  current density ${\mathbf j}= \{j^i(\mbf{x}, t)\}$ for the right and left sectors. For this purpose we start from the following  relations 
\begin{eqnarray}
    &&\rho(x,y,z,t) = \Psi^{\dagger}(x,y,z,t) \Psi(x,y,z,t) = \psi^{\dagger}(y,z) \psi(y,z),  \\
    && j^i(x,y,z,t) = v_F \overline{\Psi}(x,y,z,t) \gamma^i \Psi(x,y,z,t)= v_F \psi^{\dagger}(y,z) \alpha^i \psi(y,z),
\end{eqnarray}
where the symbol $\overline{\Psi} = \Psi^{\dagger}\gamma^0$ is the {Dirac adjoint} of $\Psi$, $\alpha_i = \gamma^0 \gamma ^i$, and the matrices $\gamma^i$ are those defined in Eq. (\ref{representation}). {Recalling the matrices $\alpha^i$ in Eq. (\ref{ALPHAMATRIX}) we can write $j^i= -v_F \psi_l^\dagger \sigma^i \psi_l + v_F \psi_r^\dagger \sigma^i \psi_r$,
which we want to read as $ j^i=j_l^i +j_r^i $.}
However, as we are interested also in chiral phenomena (given as the difference between  $l$-quantities and  $r$-quantities), we calculate separately the $l$ and $r$-parts of the probability  and current densities
\beq
 \rho_{l,r}(y,z) = \psi_{l,r}^{\dagger}(y,z) \psi_{l,r}(y,z), \qquad 
j_{l,r}^i(y,z) = {\chi v_F \psi_{l,r}^{\dagger}(y,z) \sigma^i \psi_{l,r}(y,z),}
\label{LRCURRENTS}
\eeq
where $\chi=\pm$ is the chirality label for left ($-$) and right ($+$) chirality, respectively. The explicit expressions are
\begin{eqnarray}
    && \rho_l(y,z) =  \left(|\psi_1(x,y)|^2 + |\psi_2(x,y)|^2 \right),\qquad \ \ \ \ \rho_r(y,z)= \left(|\psi_3(x,y)|^2 + |\psi_4(x,y)|^2 \right),
\label{RHOS}\\
	&& j^x_{l}(y,z) = -2v_F\text{Re} 			\left[\psi_1^{\ast}(y,z) \psi_2(y,z)\right],\qquad\ \ \ \ \ \  j^x_{r}(y,z)=2v_F\text{Re} 			\left[\psi_3^{\ast}(y,z) \psi_4(y,z)\right],
\label{JXS}\\
&&
    j^y_{l}(y,z) = -2v_F \text{Im} \left[\psi_1^{\ast}(y,z) \psi_2(y,z)\right],\qquad\ \ \ \ \ \  j^y_{r}(y,z)=2v_F \text{Im} \left[\psi_3^{\ast}(y,z) \psi_4(y,z)\right],
 \label{JYS}\\
&&    j^z_{l} (y,z)= v_F \left( -|\psi_1(y,z)|^2 + |\psi_2(y,z)|^2\right),\qquad j^z_{r }(y,z)=v_F\left( |\psi_3(y,z)|^2 - |\psi_4(y,z)|^2\right),
\label{JZS}
\end{eqnarray}
where we  used the relations (\ref{SPINORS}). 

{
The above definitions for the currents and densities, together with the general form  of the spinor $\Psi$ in (\ref{GENSPINOR21}) yield the resulting  parity properties indicated in Table \ref{yzparity}. They are obtained using the symmetry relations (\ref{Z_properties}) for the 
functions ${\cal Z}^{\pm}$ together with the well known properties of the eigenfunctions  of the one dimensional harmonic oscillator ${\cal Y}^{\pm}$. } 
\begin{table}[h]
	\centering
	{
		\renewcommand{\arraystretch}{1.5}
		\setlength{\tabcolsep}{9pt}
		\begin{tabular}{c c c }
			\hline\hline
			Bilinear & $ z$-parity & $y$-parity   \\ \hline
			$ j_l^x, \,\, j_r^x$ & $-$ & $-$  \\
			$j_l^y, \,\, j_r^y$ & $+$ & $-$  \\
			$j_l^z, \,\, j_r^z$ & $+$ & $+$  \\
			$\rho_l, \,\, \rho_r$ & $+$ & $+$  \\ 
			\hline\hline
		\end{tabular}
	}
\caption{{$z$-parity  and $y$-parity of the current and the probability densities.}}
	\label{yzparity}
\end{table}

We plot the probability density and currents for each left/right chirality [Eqs. (\ref{RHOS})-(\ref{JZS})] in Figs. \ref{probability}-\ref{currentZ}, where we use the values: $B_0=1.0\,$T, $b=0.26$\,nm$^{-1}$, $k_x=0\,$nm$^{-1}$, $L=100$nm and $n=0$, with corresponding dimensionless root energy $\varepsilon_{0,1} = 0.6188$ ($m=1$), for which the values of the constants in Eq. (\ref{DEFZRZL1}), $a_{1l}=1.0$, $a_{1r}=0.3452$, $a_{2l}=-0.5853$ and $a_{2r}=-0.5853$ are obtained. {From Fig.~\ref{probability} we observe left-right chiral charge imbalance, i.e. the axial probability density $\rho_5$ is nonzero. The probability density is well-localized in the $y$-direction, while in $z$-direction it is distributed in the bulk. Focusing now on the density currents Figs.~\ref{currentX}-\ref{currentZ}, we find nonzero charge current densities except from $j_l^x$, which is approximately (numerically) zero. Again, we find a good localization of currents in $y$-direction and a distribution over the bulk. These figures are consistent with the parity properties indicated in Table \ref{yzparity}. This can be better visualized from Fig. \ref{slices}, where cuts of the electrical quantities (left plus right quantities) at $y=0$ and $z=0$ are shown [Figs. \ref{slices}$a$) and \ref{slices}$b$), respectively]}.

{
Since we are dealing with a classical version  of the Dirac equation coupled to the electromagnetice field, (i.e. we are not including quantum electrodynamics effects), we expect the left and right  currents $j^\mu_l, j^\mu_r$ ,  to be separately  conserved. This leads  to the conservation of the probability current  $j^\mu$  as well as of the chiral current $j_5^\mu$. As shown  in the Appendix \ref{APPB}, these conservations are a direct consequence of the  spinor $\Psi$ satisfiying the coupled Dirac equation, which we have explicitly verified in the Appendix \ref{APPE} for our solution. Moreover, starting from the explicit expressions for the currents in terms of the functions ${\cal Y}^{\pm}$ and ${\cal Z}^{\pm}_{l,r}$ in Eqs. (\ref{F1})-(\ref{F3}) we calculated the divergence for the left and right currents obtaining  a null result again. Let us emphasize that this calculation can be  carried  out analytically being a manifestation of the supersymmetry of the system . The general idea is to   carefully express the derivatives in terms of the intertwining operators ${\hat L}^ \pm_\phi$, ${\hat L}^ \pm_A$ such that   the intertwining relations (\ref{INTERTWA}) and (\ref{INTERTWB})  can  be applied afterwards. We have also carried out the numerical verification of these conservation laws,  which provides a  strong consistency check of the  results we present in figures and tables.

}

\subsection{Spin dependence}
In order to gain insight about the spin-dependence of the particle's motion, we compute the spin-density and the spin current density. {In this section we set $v_F=\hbar=1$}.
Invariance of the Dirac action under Lorentz transformations induces the total angular momentum current via  Noether's theorem. The orbital and spin contributions are
\begin{eqnarray} 
&& L^{\mu\nu\lambda}= T^{\mu \lambda} x^\nu- T^{\mu\nu} x^\lambda, \qquad
 S^{\mu\nu\lambda}=\frac{1}{2}{\bar \Psi}(x) \gamma^\mu \, \sigma^{\nu\lambda} \Psi(x), 
\end{eqnarray}
respectively.  Here $T^{\nu \lambda}$ is the {canonical}  energy momentum of the Dirac field and $\sigma^{\nu \lambda}= \frac{i}{2}[\gamma^\nu, \gamma^\lambda]$. The total angular momentum current $J^{\mu\nu\lambda}=L^{\mu\nu\lambda}+ S^{\mu\nu\lambda}$ is conserved in the non-interacting case. Let us focus now in the sector $S^{\mu ij}$ with components
\beq
S^{0 ij}=\epsilon^ {kij}\frac{1}{2} \Psi^\dagger (x) \, \Sigma_k \Psi(x), \qquad  S^{p ij}=  \epsilon^ {kij}\frac{1}{2} \Psi^\dagger (x) \alpha^p \Sigma_k \Psi(x).
\eeq
In Dirac space,  $\Sigma_k$ is the spin operator and $\alpha^p$ is the velocity operator, having the form 
\begin{equation}
    \Sigma_k = \left(
    \begin{matrix}
        \sigma_k & 0 \\
        0 & \sigma_k
    \end{matrix}
    \right), \qquad 
\alpha^p = \left(
    \begin{matrix}
        -\sigma_p & 0 \\
        0 & \sigma_p
    \end{matrix}
    \right),
\end{equation}
respectively. As usual for antisymmetric indices in three dimensions we define $S^{\mu ij}=\epsilon^{ijk} S^\mu_k$, yielding a vector expression $S^\mu_k$ for the currents. Then we read
\beq
S^0_k=\frac{1}{2}\Psi^\dagger(x) \Sigma_k\Psi(x), \qquad  S^p_k  = \frac{1}{2}\Psi^\dagger(x) \alpha^p \,\Sigma_k\Psi(x).
\eeq

The spin density in $k$-direction is $S^0_k$ while the spin current density $S^p_k$ measures the spin density in the direction $k$ carried along the direction $p$ \cite{chinos}. Notice that $S^p_k$ is not real so in the following the spin current density in the symmetrized form $\left(S_k^p\right)_{\text{sym}}=\frac{1}{4} \Psi^\dagger (\alpha^p \Sigma_k + \Sigma_k \alpha^p ) \Psi$ will be used \cite{SCIREP}. Since $\alpha^p$ is the velocity operator in Dirac space, these relations follow the classical expression $J^p= \rho v^p$ associated to a density $\rho$, which in this case has indices $k$ for the three spin projections.

\subsubsection{The spin density}
It is computed by
\begin{align}
    S^0_k(y,z) = {\frac{1}{2}}\Psi^{\dagger}(y,z) \Sigma_k \Psi(y,z),
\end{align}
where the subindex $k$ refers to the direction of the spin projection. The calculation yields
\beq
 S^0_k = \frac{j_{k,r}-j_{k,l}}{2}=-\frac{j_{k,5}}{2}, 
\end{equation}
{that is, the spin densities are proportional to the corresponding axial currents previously obtained.}

\subsubsection{The spin current density }

{As mentioned before, we use the symmetric form of the spin current operator
\beq
    \left(S_k^p\right)_{\text{sym}}=\frac{1}{4}\left(\Sigma_k \alpha^{p}+\alpha^{p} \Sigma_k\right)  = \frac{1}{2}\delta^p_k \gamma_5.
\eeq
With this definition, it is straightforward to express the spin current density in terms of previously calculated currents. In fact we have
\beq
S^p_k= \frac{1}{2}\delta^p_k \Psi^\dagger \gamma_5 \Psi = \delta^p_k \rho_5.
\label{spinCurr}
\eeq
In the last expression, $\rho_5$is calculated following Eq. (\ref{RHOS}). We obtain nonzero spin density currents which reflect the chiral character of the system.

The vanishing of the cross-components in the symmetrized framework implies the strict spin-momentum locking in the bulk of Weyl semimetals, a hallmark feature reported both theoretically (Ref.~\cite{armitage_RevModPhys.90.015001}) and experimentally via spin-resolved ARPES (Ref.~\cite{lv2015observation1}).} 

{ \section{The degeneracy of the system }

\label{DEG}

Since the determinant  of the matrix (\ref{DISPREL0}) yielding the energy eigenvalues is independent of the quantum numbers  $\eta$ and $\kappa$ entering the wavefunction (\ref{GENSPINOR21}) a four-fold degeneracy appears.  We have not been able to identify commuting operators with the Hamiltonian and with themselves,  from which  such quantum numbers would arise. However we can highlight how the measuring of some bilinears of the system  can distinguish among the  different choices.

For example, let us observe that taking either $\kappa=+1$ or $\kappa=-1$ changes the relative signs of the spinor components $\psi_2$ and $\psi_4$. As a consequence, the probability density $\rho$ and the longitudinal current density $j^z$ remain unchanged, whereas the transverse current densities $j^x$ and $j^y$ acquire an overall minus sign. Thus, the direction of such currents serves to identify the corresponding  value of $\kappa$.

 On the other hand, even though there is no axial anomaly in the system, i.e. the axial current is conserved, we can still have a chiral violation due to interactions  that mix the left and right sectors. We can  look at this possibility observing the analogous of the condensate $\langle {\bar \Psi} \Psi \rangle $, which acts as a classical parameter representing the dynamical mass of the system. Recalling ${\bar \Psi} \Psi= 2\rm{Re}(\psi _{1}^{\ast }\psi _{3}+\psi _{2}^{\ast }\psi
_{4})$ we find 
\begin{equation}{\bar \Psi} \Psi= +2\eta |\mathcal{A}|^{2}\rm{Im} \left[ \mathcal{Z}_{l}^{+}(z)\,%
\mathcal{Z}_{r}^{+}(z)\,\left( \mathcal{Y}^{+}\right) ^{2}-\left[ \left( 
\mathcal{Z}_{l}^{-}(z)\,\mathcal{Z}_{r}^{-}(z)\right) \,\right] \left( 
\mathcal{Y}^{-}\right) ^{2}\right].
\end{equation}
Let us emphasize that this result is independent of $\kappa$, providing  a direct measure of the remaining quantum number  $\eta$.

}

\section{Conclusions} \label{conclusions}

{We  study the charge and spin transport properties in the bulk of a WSM slab immersed in an external magnetic field  in the framework of an effective low-energy theory that describes quasiparticles near the Weyl nodes.
Although the axial vector coupling in the effective Dirac equation for a WSM, characterized by the axion angle  $\theta$ in Eq. (\ref{DIRAC0}), can be eliminated via a chiral transformation, its effects manifest in the material's electromagnetic response, which is governed by axionic electrodynamics. Observe that even though  the parameter  $\theta$ typically includes discontinuous boundary terms that account for nearby materials, these effects are fully eliminated by the transformation. In this way, the electromagnetic  response enters into the standard Dirac-Weyl equation via the minimal coupling of electric and magnetic fields sourced by the external magnetic field. We consider an external homogeneous magnetic field in $z$-direction, which induces a linear electric field ($\mathbf{E}\propto z$) in $z$-direction, as a consequence of the magnetoelectric effect according to Eq.  (\ref{Modified_Eqs}).   In the Landau gauge, the resulting Dirac-Weyl equation can be naturally decoupled into electric and magnetic contributions. { The corresponding factorized second-order operators exhibit a supersymmetric and, for the specific electrostatic potential considered here, a PT-supersymmetric structure, respectively. We stress that the underlying Dirac-Weyl Hamiltonian coupled to real electromagnetic fields remains Hermitian throughout; the PT symmetry arises only at the level of the auxiliary operators introduced in the factorization procedure.} The identification of the corresponding intertwining operators, summarized in the Appendix \ref{INTERT},  plays a crucial role in providing an analytical verification of the consistency  of our results. The magnetic sector reduces to  the one-dimensional harmonic oscillator problem, while the electric sector involves linear combinations of  tri-confluent Heun's functions. Separation of variables is imposed on the spinor components (\ref{GENSPINOR21}) and the  solutions are exactly obtained by demanding Robin BCs, yielding no normal charge flux at the $z=\pm L$ edges. {These boundary conditions are introduced to define the bulk spectrum of the finite slab and are not intended to describe surface-localized Fermi-arc states.
}
These BC's determine the appropriate  linear combinations required in the electric sector as a consequence of Eq. (\ref{DISPREL0}), which provides the  eigenfunctions as well as the  spectrum $\{E_{n,m }  \}$ of the system. The latter  is characterized by  two labels: (i) the principal quantum number $n=0,1,2,\dots$, which  is inherited from the harmonic oscillator-like spectrum of the magnetic contribution, and (ii) the root quantum number, $m=\dots, -2, -1, 0, 1, 2 \dots$, which labels the roots of $\det M=0$ for a given $n$, where  $M$ is the matrix in Eq. (\ref{DISPREL0}). This determinant is independent of the {additional quantum numbers $\eta$  and $\kappa$, indicating  that the system shows a four-fold degeneracy labeled  by  $\eta=\pm 1$ and $\kappa=\pm 1$ in the spinor (\ref{GENSPINOR21}). We restrict ourselves to the case $\eta=1,  \kappa=+1$  in this work, and  the  energy eigenvalues are parametrized  as  $E_{n,m} =\varepsilon_{n,m}\times (\hbar v_F/L)$, where $\varepsilon_{n,m}$ is a dimensionless label. Some comments  on how to identify the labels $\eta$  and $\kappa$ are briefly discussed in section \ref{DEG}.}

{Given that our starting point is the massless Dirac-Weyl equation  minimally coupled to external electromagnetic fields, the resulting charge and  chiral currents are conserved. In the Appendix  \ref{APPF} we have analytically verified these conservations laws from our explicit solutions, providing a consistency check of the results.  In other words, our solution does not have access to the chiral anomaly, which aligns with our objective of addressing the classical issue without involving quantum electrodynamics. Additionally, from the symmetry properties of the solution we are able to determine the parity of the currents and densities, which we report in Table \ref{yzparity}.
}

{In our numerical estimations we take $v_F= 5 \times 10 ^{5}$m/s, $b=0.26 \, {\rm nm}^{-1}$ and $L=100 \, {\rm mn}$ and present different results  for the {dimensionless energies}   ${\varepsilon}_{n,m}$ [obtained from the condition det$M= Z1\times Z2=0$, Eq. (\ref{Z1Z2})] as functions of the remaining parameters. For $B_0= 1\, {\rm T}$ some values of  these energies are given in Tables \ref{tab:Evsn_Z2} and \ref{tab:Evsn_Z1} (roots of $Z2=0$ and $Z1=0$, respectively) for $n=0,1,2,3,4$. In particular, for $n=0$ the corresponding roots are shown in Fig. \ref{energies} for $B_0=1\rm{T}$ and in Fig. \ref{energiesB} for different values of $B_0$. We plot the dimensionless energies $\varepsilon_{n,m}$ as a function of $n$ for different values of the root quantum number $m$ in Fig. \ref{Evsn}, finding a separation between even- and odd-$n$ energy values which each behaves linearly with $n$. Also, for $n=0$, the dimensionless energies $\varepsilon_{n,m}$ are shown as a function of $m$ for different values of the magnetic field $B_0$ (Fig. \ref{EvsB}), where a linear behavior is found and a collapse (positive energies going to negative energies) in energy levels occurs for sufficiently large magnetic fields. These values are also shown in Tables \ref{tab:EvsB_Z2} and \ref{tab:EvsB_Z1}. }

We compute the chiral projections of the probability densities in Fig. \ref{probability}, finding chiral charge imbalance. We calculate also the corresponding chiral projections of the probability currents finding a vanishing charge current only in the $x$-direction for left chirality, i.e. in this direction that is perpendicular to the applied electric and magnetic fields, the material presents contributions to the electric/axial current just from one chirality (right). This is plotted  in Fig. \ref{currentX}.  On the other  hand, non-zero chiral  currents arise in $y$ and $z$-directions, as shown in Figs. \ref{currentY} and \ref{currentZ}, respectively. The axial current $j_5^z$ points along the direction connecting the two Weyl nodes and it is reminiscent of the net chiral exchange between the nodes that arises from the axial anomaly, via  the bending of the bands in a more complete quantum  description. {These behaviors are directly tied to novel transport effects in WSMs, specifically the planar Hall effect \cite{PHE, PHE2, pH2}
and the chiral separation effect \cite{CSE}, respectively. We recall that the planar Hall effect differs from the conventional
Hall effect in that the induced transverse current lies in the plane spanned by the electric and magnetic fields,
whereas in the ordinary Hall configuration the transverse current is perpendicular to that plane.}

{Focusing on the spin dependence of the system, we find that spin densities are proportional to the corresponding axial currents, \textit{i.e.} $S^0_k\propto j_{5}^k$, whereas spin currents are associated to electric currents through Eq. (\ref{spinCurr}).}

This study aims to shed light on the understanding and manipulation of transport phenomena in WSM's, demonstrating how the interplay between axionic response and supersymmetry governs the dynamics of quasiparticles in the material, offering novel insights into their effective field theory description.

\acknowledgments{ J.C.P.-P was supported by the SECIHTI under the program ``Estancias Posdoctorales por México'' with CVU number 671687. All authors acknowledge partial support by UNAM-PAPIIT project No. IG100224, UNAM-PAPIME project No. PE109226 and by SECIHTI project No. CBF-2025-I-1862. A.M.-R. also acknowledges financial support by the Marcos Moshinsky Foundation. We thank Ricardo Martínez for useful conversations. }

\appendix

\section{A brief on  the factorization method}

\label{FACTHAM}

We recall some basic features of this method, which we use along the manuscript. To this end let us consider the basic operators
\beq
q_\chi=p +\chi  W(x),   \qquad (q_\chi)^\dagger=p +\chi  W^*(x), \qquad \chi=\pm 1,
\eeq
where $p=-i \frac{d}{dx}$ is the one dimensional momentum operator and $W(x)$ is a complex function for the moment.  Here we set $\hbar=1$ and omit the hat to label operators. From these two operators we focus on the following two factorized Hamiltonians 
\beq
H_{(+, -)}= q_+ q_{-}, \qquad  H_{(-, +)}= q_- q_{+}.
\label{DEF}
\eeq
The relations
\beq
q_{-}H_{(+, -)}= H_{(-,+)} q_{-}, 
\qquad q_{+}H_{(-, +)}= H_{(+,-)} q_{+}
\label{FACT}
\eeq
are satisfied by construction. The operators $q_{\pm}$ are dubbed intertwining operators since they allow to relate the eigenfunctions of the Hamiltonians $H_{(+,-)}$ and $H_{(-,+)}$, as illustrated in the following. Introducing  the eigenvalue equations 
\beq
H_{(+, -)} | E_{(+, -)}\rangle= E_{(+, -)}| E_{(+, -)}\rangle, \qquad 
H_{(-, +)} | E_{(-, +)}\rangle= E_{(-, +)}| E_{(-, +)}\rangle, 
\eeq
the relations (\ref{FACT}) yield
\beq
q_{-}|E_{(+, -)} \rangle= \lambda_{(-, +)}
|E_{(-,+)} \rangle, \qquad 
q_{+}|E_{(-, +)} \rangle= \lambda_{(+, -)}
|E_{(+,-)} \rangle, 
\label{INTER}
\eeq
provided there is no degeneracy. This means that $ H_{(+,-)}$ and $H_{(-,+)}$ share the eigenvalues $E_{(+,-)}= E_{(-,+)} \equiv E_+= E_{-}= E$ with their eigenfunctions related by the intertwining operators $q_\chi$. The explicit form of the Hamiltonians are
\beq
H_{(+, -)}= p^2-W^2+  i \frac{d W(x)}{dx}, \qquad H_{(-,+)}= p^2-W^2-  i \frac{d W(x)}{dx}.
\eeq
From here we distinguish  the two cases arising in the manuscript: (1)  $W(x)= - i w(x)$  and (2) $W(x)=  w(x) $,  with $w(x)$ a real function. We separately consider each case.

\subsection{Case \texorpdfstring{$W(x)=-i w(x)$}{W(x)=-i w(x)}}
In a compact notation we have 
\begin{eqnarray}
&& H^{(1)}_{(\chi, -\chi)}= p^2+ w^2+ \chi  \frac{d w(x)}{dx},
\end{eqnarray}
yielding an hermitian Hamiltonian. Besides 
\beq
q_\chi=p - i \chi w(x), \qquad (q_\chi)^\dagger= q_{-\chi},
\eeq
such that $H^{(1)}_{(\chi, -\chi)}$ are hermitian and positive definite. This is the case of standard supersymmetric quantum mechanics in one dimension and the Hamiltonians $H^{(1)}_{(+, -)}$ and $H^{(1)}_{(-, + )}$ are called supersymmetric partners. The term $w(x)$ is dubbed the superpotential.

Let us recall the main properties of these systems. The Hamiltonians can be written in the manifest hermitian form $H_{(+, -)}= q_+ q_-= q_+ (q_{+})^\dagger$ and $H_{(-, +)}= q_- q_+= (q_+)^\dagger (q_{+})$ in term of the intertwining operator $q_+$. 

The relations (\ref{INTER}) imply hat we can take 
\beq
\lambda_{(+, -)}= \lambda_{(-, +)} = \sqrt{E} , 
\label{LAMBDAREL}
\eeq  
up to a phase. 

Now let us consider the ground state of the supersymmetric system, which corresponds to zero energy. We start by demanding 
$q_+ q_+^\dagger|0_{(+,-)}\rangle=0$ and   
$ q_+^\dagger q_+|0_{(-,+)}\rangle=0$, which in turn require $q_+^\dagger|0_{(+,-)}\rangle=0$ and $q_+|0_{(-,+)} \rangle$, respectively. However, we next show that both conditions cannot hold simultaneously. To this end we solve the corresponding differential equations
\beq
\frac{1}{i} \frac{d \Psi_{0+,-}}{dx}+iw(x)\Psi_{0+,-}=0, \qquad 
\frac{1}{i} \frac{d \Psi_{0 -,+}}{dx}-iw(x)\Psi_{0 -,+}=0,
\eeq
which are summarized in 
\beq
\frac{d \Psi_{ \sigma}}{dx}=+\sigma \frac{d u(x)}{dx} \Psi_{ \sigma}, \qquad \sigma =\pm 1,
\eeq 
with the choices $\Psi_{\sigma=1},\,(\Psi_{\sigma=-1})$  for $\Psi_{0+,-},\, (\Psi_{0-,+})$, respectively. We have written
$w(x)=\frac{d u(x)}{dx}$ for clarity. The solution is 
\beq
\Psi_{ \sigma}(x)= \Psi_{ \sigma}(0)\, \exp\ \left\lbrace\sigma \Big(u(x) -u(0)\Big)\right\rbrace.
\eeq
For convergence we have to demand $\lim_{x \rightarrow \pm \infty}\Psi_{ \sigma}(x)=0 $, which means that $\lim_{x \rightarrow \pm \infty}u(x)= \pm \infty$ according to the sign of $\sigma$. Then we realize that when we manage to fulfill this condition for a given $\sigma$, the solution $\Psi_{-\sigma}(x)$ becomes non-normalizable and must be excluded. In this way, there is only one normalizable solution for the zero-energy ground state.

\subsection{Case \texorpdfstring{$W(x)={\tilde w}(x)$}{W(x)={\tilde w}(x)}}
Here we have 
\begin{eqnarray}
&& H^{(2)}_{(\chi, -\chi)}= p^2-{\tilde w}^2+ i \chi  \frac{d {\tilde w}(x)}{dx},
\end{eqnarray} 
resulting in a non-hermitian Hamiltonian, which follows into the category of PT-symmetric Hamiltonians \cite{BENDER}. Their relevance arises because they can have real eigenvalues in spite of being non-hermitian, though this is not a general property.

In this case we have $q_\chi= (q_\chi)^\dagger$ which implies $H^{(2)}_{(+,-)}= (H^{(2)}_{(-,+)})^\dagger$. 
Nevertheless, what is important for our purposes is that the defining factorization (\ref{DEF}) preserves the intertwining relations (\ref{INTER}) that allow to relate the eigenfunctions of $H^{(2)}_{+,-}$ and $H^{(2)}_{-,+}$. That is to say, these Hamiltonians  still share a common eigenenergy $E_{(+,-)}=E_{(-,+)}= E$.
The intertwining relations (\ref{INTER}) also yield the weaker condition
\beq
\lambda_{(+,-)}\, \lambda_{(-,+)}= E,
\eeq
as the analogous of Eq. (\ref{LAMBDAREL}). In an abuse of language we call ${\tilde w}$ the superpotential in the case of a PT-symmetric Hamiltonian.  Notice that we can go from the supersymmetric partners 
$H^{(1)}_{(\chi, -\chi)}$ to what we call  the PT-supersymmetric partners  $H^{(2)}_{(\chi, -\chi)}$ by taking an imaginary superpotential $w=i{\tilde w}$.
The extension of standard supersymmetry to PT-supersymmetry  has gained much attention in the literature \cite{DAS} and its further consequences are beyond the scope of this work. For our purposes it is enough to take into account the intertwining relations (\ref{INTER}).

\section{Basic properties  of the Dirac equation}
\label{APPB}
 Following the notation
 \beq
 x^0=v_F t, \qquad  p_\mu=i\hbar \partial_\mu, \qquad A_\mu=( \phi / v _{F} , \mbf{A}), \qquad \gamma^\mu=(\gamma^0, \boldsymbol{\gamma}) , 
 \eeq
 we can write the Dirac equation coupled to a real external electromagetic field given in (\ref{EDIRAC}) as 
 \beq
 i\hbar \gamma^\mu \partial_\mu  \Psi= e A_\mu \gamma^\mu \Psi.
 \label{DIRAC1}
 \eeq
Taking the adjoint and recalling that
\beq
\gamma^0 (\gamma^\mu)^\dagger \gamma^0=\gamma^\mu, \qquad {\bar \Psi=\Psi^\dagger \gamma^0} , 
\eeq
we obtain
\beq
i\hbar \partial_\mu {\bar \Psi} \gamma^\mu = - e A_\mu {\bar \Psi} \gamma^\mu.
\label{DIRAC2}
\eeq
From Eqs. (\ref{DIRAC1}) and (\ref{DIRAC2}) it is a direct calculation to show that
\beq
\partial_\mu ({\bar \Psi} \gamma^\mu \Psi)=0, \qquad 
\partial_\mu ({\bar \Psi} \gamma^5 \gamma^\mu \Psi)=0,
\eeq
at the classical level (no anomalies), indicating the conservation of  both the charge current as well as the chiral current, respectively.
Since in our case chirality is a good quantum number, this means that the left and right handed currents in (\ref{LRCURRENTS}) are also separately conserved.

\section{ Robin (Mixed) Boundary conditions in this system}

\label{NORM2}

{The Dirac operator ${\hat D}$, being of first order in the derivatives requires Robin boundary conditions (BCs) when the space time  ${\cal M}$ has boundaries $\partial {\cal M}$. These were first obtained in Ref. \cite{BRANSON1992249} and we follow the version discussed in Ref. \cite{astaneh2023fermions} adapted for the $3+1$ dimensional case. The basic idea is to separate the Dirac spinor $\Psi$ into halfs  and apply Dirichlet BC  over one half of the components together with derivative BC on the remaining half. The partition of the spinor components is achieved  by chiral projectors $\Pi_{\pm}$
\begin{equation}
\label{projectors}
    \Pi_{(\pm, {\hat{n}})} = \frac{1}{2}\left(\mathbb{I}\pm \Gamma_{\hat n }\right),
\end{equation}
where $\Gamma_{\hat n}=\Gamma^5\gamma^n$ and   $\gamma^n=n_\mu \gamma^\mu$, where $n_\mu$ is the exterior normal vector at the interface. The additional property $\Pi_{(+, {\hat n} )}\gamma^n= \gamma^n \Pi_{(-, {\hat n})}$ is required, which demands $\{ \Gamma^5, \gamma^n \}=0$ that can be realized by choosing $\Gamma^5=\gamma^5$. In the case of spacelike boundaries  $\gamma^n= n_i \gamma^i$, where $(\gamma^i)^\dagger=-\gamma^i$ and $(\gamma^5)^\dagger= \gamma^5$ in our conventions. This makes the projectors hermitian: $[\Pi_{\pm}]^\dagger=\Pi_{\pm}$
We recall the explicit expression of our gamma matrices
 [Eqs. (\ref{representation})]:
\begin{equation}
    \gamma^0= 
    \begin{pmatrix}
    0 & \sigma_0\\
    \sigma_0 &0
    \end{pmatrix},\qquad \gamma^i= 
    \begin{pmatrix}
    0 & \sigma^i\\
    -\sigma^i &0
    \end{pmatrix},\qquad \gamma^5= 
    \begin{pmatrix}
    -\sigma_0 & 0\\
    0 & \sigma_0
    \end{pmatrix}.\ 
\end{equation}

In terms of these elements, the boundary conditions are 
\beq
\Pi_{(-,{\hat n}) }\Psi |_{\partial{\cal M}}=0, \qquad \partial_n \Pi_{(+, {\hat n} )}\Psi |_{\partial{\cal M}}=0.
\label{BCS}
\eeq
In order to dilucidate the meaning of such BC we restrict  to our  problem where we have edges at $z=\pm L$, with their normal vectors being $ + \hat{\mathbf{k}}$  and  $ - \hat{\mathbf{k}}$, respectively, which  yields    $\gamma^{ n}=\pm \gamma^3$ for each border. Then
\begin{equation}\label{chi}
    \Gamma_{\pm \hat{\mathbf k}}=\left\lbrace \begin{matrix}
    \gamma^5\gamma^3,\ \text{for }z=+L,\\
    -\gamma^5\gamma^3,\ \text{for }z=-L,
    \end{matrix}
    \right.\qquad \text{with } \gamma^5\gamma^3=\begin{pmatrix}
        0 & -\sigma^3\\
        \sigma^3 & 0
    \end{pmatrix}.
\end{equation}
To avoid confusion we define the operators
\begin{equation}
    \Pi_{\pm} = \frac{1}{2} \left(\mathbb{I}\pm \gamma^5\gamma^3 \right) = \begin{pmatrix}
        1 & 0 & \mp 1 & 0\\
        0 & 1 & 0 & \pm 1\\
        \mp 1 & 0 & 1 & 0\\
        0 & \pm 1 & 0 & 1
    \end{pmatrix},
\end{equation}
in terms of which we can write all the required directional projectors $\Pi_{(\pm,{\hat {\mathbf k}}) }$. In fact we have
\beq
\Pi_{(-,{\hat {\mathbf k}}) }= \Pi_- , \qquad 
\Pi_{(-,{-\hat {\mathbf k}}) }= \Pi_+ , \qquad 
\Pi_{(+,{\hat {\mathbf k}}) }= \Pi_+ , \qquad 
\Pi_{(+,{-\hat {\mathbf k}}) }= \Pi_- . \qquad 
\label{DETPROJ}
\eeq
The above equations indicate  that the chiral projectors are interchanged at the $z=\pm L$ edges, as can be seen from Eqs. (\ref{projectors}) and (\ref{chi}), after the change $\hat{\mathbf{k}}$ into  $-\hat{\mathbf{k}}$. 

{Next we illustrate the meaning of the BCs (\ref{BCS}) taking $z=+L, {\hat n}= \hat {\mathbf k} $ as the corresponding boundary, with the required projectors given in Eq.(\ref{DETPROJ}).
The first BC in (\ref{BCS}) ensures that the current flowing perpendicularly to the boundary is zero there.  We have
\beq
J_z|_{z=L}=({\bar \Psi} \gamma^3 \Psi)|_{z=L}=
({\bar \Psi}(\Pi_- +\Pi_+) \gamma^3 \Psi)|_{z=L}= 
({\bar \Psi}(  \gamma^3 \Pi_- \Psi)|_{z=L}=0 . 
\eeq
In the last step we used the hermiticity of the projector $\Pi_-$ together with the commutator $[\gamma^0, \Gamma]=0$ such that 
${\bar \Psi} \Pi_- = \Psi^\dagger \Pi_-^\dagger \gamma^0=0$ at the boundary. The remaining projector $\Pi_+$ is moved to the right using $ \Pi_+\gamma^3=\gamma^3 \Pi_- $, yielding the desired result. Similar result follows for the boundary at $z=-L$.

The second BC in (\ref{BCS}) results from the application of the projector $\Pi_-$ to the full Dirac equation (\ref{DIRAC1}) at the boundary. We have 
\beq
 \Pi_-\Big(i\hbar \gamma^\mu \partial_\mu  \Psi\Big)= e \Pi_- \Big( A_\mu \gamma^\mu \Psi\Big)= e A_\mu  \gamma^\mu  \Pi_- \Psi=0.
 \label{BCS2}
 \eeq
Since the non-zero potentials in our problem are $A_0$ and $A_x$,  we can move $\Pi_-$ to the right across $A_\mu \gamma^\mu$ because 
$\gamma^0$ and $\gamma^1$ commute with $\Pi_-$. Still we are left with the left-hand side of (\ref{BCS2}), which we now verify that yields the  BC $\partial_3\Pi_+ \Psi|_{\partial{\cal M}}=0$ in (\ref{BCS}). Again, we can move $\Pi_-$ to the right across $(\gamma^0 \partial_0+ \gamma^1 \partial_1+\gamma^2 \partial_2) $ to get a zero contribution, yielding 
\beq
0=i \hbar \, \Pi_- \gamma^3 \partial_3 \Psi= i \hbar \,  \gamma^3  \Pi_+\partial_3 \Psi= i \hbar \,  \gamma^3  \partial_3 \Pi_+ \Psi.
\label{BCS3}
\eeq
Since $(\gamma^3)^2=-1$ we recover  the second BC in (\ref{BCS}).
 To simplify notation we have not indicated explicitly that all the above relations in (\ref{BCS2}) and (\ref{BCS3}) are to be evaluated  at the boundary.
 }

\subsection{{The Dirichlet boundary condition}}

\label{APPC11}

Taking care of the directions of the normal vector at each boundary,  the conditions $\Pi_{(-, {\hat n})}\Psi| _{\partial {\cal M}}$ read
\begin{equation}\label{Dirichlet}
    \Pi_-\Psi(x,y,z)\Big{|}_{z=+L}=0, \qquad \Pi_+\Psi(x,y,z)\Big{|}_{z=-L}=0.
\end{equation}
Conditions in Eq. (\ref{Dirichlet}) reduces to
\begin{eqnarray}
        && \psi_1|_{z=+ L}=-\psi_3|_{z=+ L},\qquad
        \psi_2|_{z=+ L}=\psi_4|_{z=+ L},
        \label{condition1}\\
        && \psi_1|_{z=- L}=\psi_3|_{z=- L},\qquad \quad 
        \psi_2|_{z=- L}=-\psi_4|_{z=- L}.
        \label{condition2}
\end{eqnarray}

Recalling the form of the spinor
\begin{equation}
    \begin{pmatrix}
        \psi_1\\
        \psi_2\\
        \psi_3\\
        \psi_4
    \end{pmatrix} = e^{ik_x x}\begin{pmatrix}
        \mathcal{A}\mathcal{Z}_{l}^+(z) \mathcal{Y}^-(y)\\
        \mathcal{B}\mathcal{Z}_l^-(z) \mathcal{Y}^+(y)\\
        \mathcal{C}\mathcal{Z}_r^-(z) \mathcal{Y}^-(y)\\
        \mathcal{D}\mathcal{Z}_r^+(z) \mathcal{Y}^+(y)
    \end{pmatrix} ,
    \label{GENSPINOR}
\end{equation}
where $\mathcal{A},\mathcal{B},\mathcal{C},\mathcal{D}$ are constants to be determined, we see that the conditions (\ref{condition1}) and (\ref{condition2}) imply
\begin{subequations}
\begin{equation}
\label{Cmas}
    \mathcal{Z}_l^+(L) = -\frac{\mathcal{C}}{\mathcal{A}}\mathcal{Z}_r^-(L), \qquad 
    \mathcal{Z}_l^+(-L) = \frac{\mathcal{C}}{\mathcal{A}}\mathcal{Z}_r^-(-L),
\end{equation}
\begin{equation}
\label{Cmenos}
  \mathcal{Z}_l^-(L) = \frac{\mathcal{D}}{\mathcal{B}}\mathcal{Z}_r^+(L)  ,\qquad
   \mathcal{Z}_l^-(-L) = -\frac{\mathcal{D}}{\mathcal{B}}\mathcal{Z}_r^+(-L).
\end{equation}
\end{subequations}
The following properties 
\begin{equation}
    \mathcal{Z}^{\pm}(z)=\mathcal{Z}^{\mp}(-z)=\left[\mathcal{Z}^{\pm}(-z)\right]^{\ast},
    \label{MORERELS}
\end{equation}
which are  valid separately for  ${\cal Z}_l$ or  ${\cal Z}_r$, allows us to rewrite all the boundary conditions at $z=L$, besides obtaining restrictions  over the constants.
Let us focus on the second equation in Eqs. (\ref{Cmas}) and change $L \rightarrow -L$ using the first relation in Eq. (\ref{MORERELS}). 
{We get 
\beq
\mathcal{Z}_l^-(L)=
\frac{\mathcal{C}}{\mathcal{A}}\mathcal{Z}_r^+(L)
\label{C11}
\eeq} 
Inserting this in the right hand side of the first equation in  (\ref{Cmas})  we obtain  the condition $ \frac{\mathcal{A}^2}{\mathcal{C}^2} = -1$. Following similar steps with Eq. (\ref{Cmenos}) the analogous condition $\frac{\mathcal{B}^2}{\mathcal{D}^2} = -1$ arises. Finally, comparing  Eq. (\ref{C11}) with the first equation in (\ref{Cmenos})  we get $\frac{\mathcal{A}}{\mathcal{C}}=\frac{\mathcal{B}}{\mathcal{D}}$ . Summarizing, the conditions for the spinor coefficients are 
\begin{equation}
    \frac{\mathcal{A}^2}{\mathcal{C}^2} = -1,\qquad 
    \frac{\mathcal{B}^2}{\mathcal{D}^2} = -1,\qquad \frac{\mathcal{A}}{\mathcal{C}}=\frac{\mathcal{B}}{\mathcal{D}}.
\end{equation}
We have two cases
\beq
\mathcal{C}=i\eta\mathcal{A}, \qquad  \mathcal{D}=i\eta\mathcal{B}, \qquad \eta=\pm 1.
\label{FINALCOND}
\eeq
 At this stage we are left only with the boundary conditions at $z=+L$, which are written in the first equations of (\ref{Cmas}) and (\ref{Cmenos}). {Recalling Eq. (\ref{MORERELS}) yields ${\cal Z}^-(z)=[{\cal Z}^+(z)]^*$}  and writing these relations in terms of functions ${\cal Z}^+$ only we find that both conditions are equivalent,  yielding the final half of the boundary conditions 
\beq
\Big({\cal Z}_l^+ + i\eta \, [{\cal Z}_r^+]^*\Big)_{z=+L}=0 .
\label{FINALDIRICHLET}
\eeq
\subsection{{The derivative boundary condition}}

\label{APPC2}

Now we deal with $\partial_n \Pi_{(+, {\hat n} )}\Psi |_{\partial{\cal M}}=0$ which reduces to  $\partial_z \Pi_{+, }\Psi=0$ at $z=L$ and $\partial_z \Pi_{-, }\Psi =0$ at $z=-L$. In terms of the spinor components we have
\begin{eqnarray}
       && \partial_z\psi_1|_{z=+ L}= \partial_z\psi_3|_{z=+ L},\qquad \quad\partial_z\psi_2|_{z=+ L}= -\partial_z\psi_4|_{z=+ L},
       \label{condition3} \\
&&\partial_z\psi_1|_{z=- L}= -\partial_z\psi_3|_{z=- L},\qquad \partial_z\psi_2|_{z=- L}= \partial_z\psi_4|_{z=- L}.
\label{condition4}
\end{eqnarray}
Comparing the above equations with (\ref{condition1}) and 
(\ref{condition2}) it is evident that the derivative $\partial_z$ appears everywhere and more importantly that there is a relative change of sign between the components $\Psi_1, \Psi_3$ and $\Psi_2, \Psi_4$. Going through a detailed analysis as in the previous section  the final derivative BCs are 
\beq
\Big(\partial_z{\cal Z}_l^+ - i\eta \,  [\partial_z{\cal Z}_r^+]^*\Big)_{z=+L}=0 . 
\label{FINALDERIVATIVE}
\eeq

\section{Solving the boundary conditions}
\label{APPD}

Summarizing, Eqs. (\ref{FINALDIRICHLET}) and (\ref{FINALDERIVATIVE}) provide four real conditions that will serve us to determine  ${\cal Z}^+_{l,r}$, which are eigenfunctions of the Hamiltonian $ {\hat H}^+_\phi$, according to Eq. (\ref{EIGENEQS}), and   satisfy the symmetry properties indicated in Eq. (\ref{Z_properties}). The linearly independent solutions of the corresponding Schr\"{o}dinger equation are the functions $u_1^+$ and $u_2^+$, given in Eqs. (\ref{U1}) and (\ref{U2}) with symmetry properties written in Eq. (\ref{u_properties}). Through this process, we will also determine the dispersion relation of the system, which provides the eigenvalues of the system's energy.

To begin with, we must expand ${\cal Z}^+_{l,r}$ in terms of $u^+_{1,2}$. An important property of this expansion is that the coefficients are real, as we now show for an arbitrary ${\cal Z}^+$. Let us consider 
\beq
{\cal Z}^{+}(z)= \alpha  \, u_1(z) +  \beta\,  u_2(z).
\eeq
Since $[{\cal Z}^+(z)]^*= {\cal Z}^+(-z) $ we must have 
\beq
\alpha^{\ast}  \, [u_1(z)]^{\ast} + \beta^{\ast}  \, [u_2(z)]^{\ast}  = \alpha  \, u_1(-z) +  \beta\,  u_2(-z)= 
\alpha  \, [u_1(z)]^{\ast} +  \beta\,  [u_2(z)]^{\ast},
\eeq
where we used $u_{1,2}(-z)= [u_{1,2}(z)]^*$ in the last step.  We conclude that  $\alpha=\alpha^*$ and $\beta=\beta^*$. 

To continue we write
\beq
{\cal Z}^+_l(z)=a_{1l}\, u^+_1(z)+ a_{2l}\, u^+_2(z), \qquad 
{\cal Z}^+_r(z)=a_{1r}\, u^+_1(z)+ a_{2r}\, u^+_2(z),
\label{DEFZRZL}
\eeq
where the coefficients $a_{1l}, \, a_{1r},\, a_{2l},\, a_{2r}$ are real numbers. After substituting these expressions in the BCs (\ref{FINALDIRICHLET}), (\ref{FINALDERIVATIVE}), and separating the real and imaginary parts in each realtion   we arrive at four real  equations for the undetermined coefficients in Eq. (\ref{DEFZRZL}). We present these equations in matrix form
\begin{equation}
\begin{pmatrix}
R_1 & \eta I_1 & R_2 & \eta I_2\\
I_1 & \eta R_1 & I_2 & \eta R_2\\
R'_1 & -\eta I'_1 & R'_2& -\eta I'_2\\
I'_1 & -\eta  R'_1 & I'_2& -\eta R'_2
\end{pmatrix}
\begin{pmatrix}
        a_{1l}\\
        a_{1r}\\
        a_{2l}\\
        a_{2r}
    \end{pmatrix} =0,
    \label{DISPREL}
\end{equation}
with the notation
\beq
u_k(L)=R_k+i I_k, \qquad [\partial_z u_k(z)]_{z=L}= R'_k+i I'_k, \qquad k=1,2 , 
\eeq
which amounts to the separation of all the functions involved into real an imaginary parts. The spectrum of the system is determined by demanding the determinant of the matrix  $M$ in Eq. (\ref{DISPREL}) to be zero. We find that $\det M =  Z1 \times Z2$ with
\beq
Z1= (I_2-R_2)(R'_1+I'_1)-(I_1-R_1)(R'_2+I'_2), \qquad 
Z2= (I_2+R_2)(R'_1-I'_1)-(I_1+R_1)(R'_2-I'_2). 
\label{Z1Z2}
\eeq
This determinant turns out to be independent of $\eta=\pm 1$, while the eigenvectors will be $\eta$-dependent, leading in general to a two-fold degeneracy in the system.
For each eigenvalue  the $a$'s coefficients  are solved in terms of one of them. In other words, Eq. (\ref{DISPREL}) fixes the $a$'s coefficients up to an arbitrary  factor. Having determined the functions $Z^+_{l,r}(z)$,  the remaining functions can be obtained  by  the relation $Z^-_{l,r}(z)= {\cal Z}^+_{l,r}(-z)$, which still depend on this arbitrary factor. This arbitrariness can be incorporated by redefining    the  constant ${\cal A}$ in Eq. (\ref{GENSPINOR21}) such that at  the end of the day  our spinor solution  has  only one free constant that is determined the normalization in  Eq.(\ref{NORMALIZATION}).

\section{Verifying the Dirac equation } 
\label{APPE}
Following the standard procedure we have decoupled the Dirac equation by going to a second order system of equations, which normally has more solutions than desired. To take care of these possible ambiguities  we better substitute our solution 
\begin{equation}
    \begin{pmatrix}
        \psi_1\\
        \psi_2\\
        \psi_3\\
        \psi_4
    \end{pmatrix} = e ^{i E t/\hbar}  e^{ik_x x}\begin{pmatrix}
        \mathcal{A}\mathcal{Z}_{l}^+(z) \mathcal{Y}^-(y)\\
        \mathcal{B}\mathcal{Z}_l^-(z) \mathcal{Y}^+(y)\\
        i \eta \,\mathcal{ A}\mathcal{Z}_r^-(z) \mathcal{Y}^-(y)\\
        i \eta \, \mathcal{B}\mathcal{Z}_r^+(z) \mathcal{Y}^+(y)
    \end{pmatrix} ,
    \label{GENSPINOR1}
\end{equation}
in the Dirac equation (\ref{DIRAC1}) and check for consistency.

In this Appendix we work in units such that $\hbar=1=v_F$.  Recalling the relations $\alpha^k=\gamma^0 \gamma^k$, $\partial_x=ik_x$ and $\partial_t=iE$ the Dirac  equation (\ref{DIRAC1}) reduces to 
\beq
i\partial _{y}\alpha ^{y}\Psi +i\partial _{z}\alpha ^{z}\Psi  =\left( k_{x}+eA_{x}\right) \alpha ^{x}\Psi +\left( E+e\phi \right) \Psi 
\label{DIRAC3}
\eeq
with 
\begin{equation}
\alpha ^{x}=\left( 
\begin{array}{cccc}
0 & -1 & 0 & 0 \\ 
-1 & 0 & 0 & 0 \\ 
0 & 0 & 0 & 1 \\ 
0 & 0 & 1 & 0%
\end{array}%
\right) ,\qquad \alpha ^{y}=\left( 
\begin{array}{cccc}
0 & i & 0 & 0 \\ 
-i & 0 & 0 & 0 \\ 
0 & 0 & 0 & -i \\ 
0 & 0 & i & 0%
\end{array}%
\right) ,\qquad \alpha ^{z}=\left( 
\begin{array}{cccc}
-1 & 0 & 0 & 0 \\ 
0 & 1 & 0 & 0 \\ 
0 & 0 & 1 & 0 \\ 
0 & 0 & 0 & -1%
\end{array}%
\right) .
\label{ALPHAMATRIX}
\end{equation}%
In terms of the spinor components Eq. (\ref{DIRAC3}) results in the following four relations

\begin{eqnarray}
&&\partial _{y}\psi _{2}+ i\partial _{z}\psi
_{1} =\left( k_{x}+eA_{x}\right) \psi _{2} - \left(
E+e\phi \right) \psi _{1} \label{APPE1}\\
&&\partial _{y} \psi _{1} +i\partial _{z}\psi _{2}
=-\left( k_{x}+eA_{x}\right) \psi _{1} +\left( E+e\phi \right)
\psi _{2}  \label{APPE2} \\
&& \partial _{y}\psi _{4} + i \partial _{z}\psi _{3} 
=\left( k_{x}+eA_{x}\right) \psi _{4} +\left( E+e\phi
\right) \psi _{3} \label{APPE3} \\
&&\partial _{y}\psi _{3} + i \partial _{z}\psi
_{4}  =-\left( k_{x}+eA_{x}\right) \psi_3 -\left(
E+e\phi \right) \psi _{4} \label{APPE4}
\end{eqnarray}%
Our expectation is that the above equations produce a mixing of the  constants ${\cal A}$ and ${\cal B}$ which allows for  a consistent relation among each other. Verifiying this in each of the four equations is a strong consistency check of our calculations. We start with Eq. (\ref{APPE1}) by substituting the explicit forms of $\Psi_1$ and  $\Psi_2$
\begin{equation}
\mathcal{B}\mathcal{Z}_{l}^{-}(z) \partial _{y}\mathcal{Y}%
^{+}(y) +  \mathcal{A} \mathcal{Y}^{-}(y)i \partial _{z}\mathcal{Z}%
_{l}^{+}(z)  =   \mathcal{B} W(y) \mathcal{Z}_{l}^{-}(z)%
\mathcal{Y}^{+}(y) - \mathcal{A} U(z) \mathcal{%
Z}_{l}^{+}(z)\mathcal{Y}^{-}(y) 
\label{APPE11}
\end{equation}%
with the notation 
\begin{eqnarray}
 && W(y)=k_x+A_x(y), \qquad U(z)= E +e \phi(z)   
\end{eqnarray}
Next we calculate $\partial _{y}\mathcal{Y}%
^{+}(y) $ and $\partial _{z}\mathcal{Z}%
_{L}^{+}(z)$ in terms of judicious choice of  intertwining operators, designed to match the coordinate dependence of the functions in the right-hand side. From Eqs. (\ref{Ladder_Operators})  we choose
\beq
\partial_y=-{\hat L}_A^+ + W(y), \qquad \partial_z= -i ({\hat L}_\phi^--U(z))
\eeq
From Eqs. (\ref{INTERTWA}) and  (\ref{INTERTWB}) we read 
\beq
\partial_y {\cal Y}^+(y)= -\sqrt{E_A} \, {\cal Y}^-(y)+ W(y) {\cal Y}^+(y), \qquad i \partial_z {\cal Z}_l^+(z)=-i\sqrt{E_A} {\cal Z}_l^-(z) -U(z) {\cal Z}_l^+(z).
\eeq
Substituting in (\ref{APPE11}) we discover that the right-hand side exactly cancels with the contributions independent of $\sqrt{E_A}$ coming from the left-hand side . We are left with 
\beq
\sqrt{E_A} \,  {\cal Z}_l^-(z)\, {\cal Y}^-(y) ({\cal B}+ i {\cal A})=0
\eeq
fixing ${\cal B}= - i {\cal A}$, which leaves ${\cal A}$ as the only undetermined constant in the solution, which can be subsequently fixed by normalization. {We have verified that the remaining Eqs. (\ref{APPE2})-(\ref{APPE4}) produce the same condition.}

{

\section{Another look at current conservation} 
\label{APPF}

In this Appendix we prove current conservation $\partial_\mu (\Psi^\dagger \alpha^\mu \Psi)=0$ from the explicit expression for the current. Let us recall  the current components 
\begin{eqnarray}
&&j^x=- 2\mathrm{Re}(\psi_1^*\, \psi_2)  +2\mathrm{Re}(\psi_3^*\, \psi_4) \label{F1}\\
&&j^y= - 2\mathrm{Im}(\psi_1^*\, \psi_2) +
2\mathrm{Im}(\psi_3^*\, \psi_4)  \label{F2}\\
&&j^z= |\psi_2|^2 - |\psi_1|^2+|\psi_3|^2-|\psi_4|^2, \label{F3}
\end{eqnarray}
which are independent of $t$ and $x$. Then we only need to deal with $\partial_y j^y$ and $\partial_z j^z$. Here it is crucial to take into account the condition ${\cal B}=-i{\cal A}$ derived
in the previous section. The final form of the spinor, previous to normalization, is then 
\begin{equation}
    \begin{pmatrix}
        \psi_1\\
        \psi_2\\
        \psi_3\\
        \psi_4
    \end{pmatrix} = \mathcal{A} \, e ^{i E t/\hbar}  e^{ik_x x}\begin{pmatrix}
        \mathcal{Z}_{l}^+(z) \mathcal{Y}^-(y)\\
        -i\mathcal{Z}_l^-(z) \mathcal{Y}^+(y)\\
        i \eta \,\mathcal{Z}_r^-(z) \mathcal{Y}^-(y)\\
         \eta \, \mathcal{Z}_r^+(z) \mathcal{Y}^+(y)
    \end{pmatrix}.
    \label{GENSPINOR2}
\end{equation}
In the following steps we make ample use of the symmetry properties (\ref{Z_properties}) without giving the explicit reference in most cases. Also notice that  the currents  are independent of the explicit parameter $\eta=\pm 1$ in Eq. (\ref{GENSPINOR2}), though the functions ${\cal Z}^{+.-}_{l,r}$ will have an $\eta$ dependence. 

The relevant currents are 
\begin{eqnarray}
j^{y}&&=2 |\mathcal{A}|^{2} \mathcal{Y}^{-}\mathcal{Y}^{+}\, \mathrm{Im}\Big( i \Big(\mathcal{Z}_{l}^{+}\,(\mathcal{Z}_{l}^{-})^{\ast }-%
\mathcal{Z}_{r}^{-}\,(\mathcal{Z}_{r}^{+})^{\ast }\Big) \Big)=  2 |\mathcal{A}|^{2} \,  \mathcal{Y}^{-}\mathcal{Y}^{+}\mathrm{Im}\Big(i \Big((\mathcal{Z}_{l}^{+})^2-%
(\mathcal{Z}_{r}^{-})^2 \Big)\Big) \nonumber 
\\
j^{y} &&= 
2 |\mathcal{A}|^{2} \,  \mathcal{Y}^{-}\mathcal{Y}^{+}\mathrm{Re}\Big( \Big((\mathcal{Z}_{l}^{+})^2-%
(\mathcal{Z}_{r}^{-})^2 \Big)\Big),
\end{eqnarray}%
\begin{eqnarray}
j^{z}&&=|\mathcal{A}|^{2} \Big[(\mathcal{Y}^{-})^{2}\Big(|\mathcal{Z}_{r}^{-}|^{2}-|%
\mathcal{Z}_{l}^{+}|^{2}\Big)+(\mathcal{Y}^{+})^{2}\Big(|%
\mathcal{Z}_{l}^{-}|^{2}-|\mathcal{Z}_{r}^{+}|^{2}\Big)\Big] \nonumber  \\
j^{z} &&=|\mathcal{A}|^{2} 
\Big(\mathcal{Z}_{r}^{-} \mathcal{Z}_{r}^{+}
- \mathcal{Z}_{l}^{-} \mathcal{Z}_{l}^{+}\Big)
\Big[(\mathcal{Y}^{-})^{2}- (\mathcal{Y}^{+})^{2}\Big]  
\end{eqnarray}%
Since ${\cal Z}_r$ and ${\cal Z}_l$ are independent functions  we separate the currents in their chiral contributions
\begin{eqnarray}
&&j^y_l= 2 |\mathcal{A}|^{2} \,  \mathcal{Y}^{-}\mathcal{Y}^{+}\mathrm{Re}\Big( (\mathcal{Z}_{l}^{+})^2 \Big)  , \qquad  
j^y_r= -2 |\mathcal{A}|^{2} \,  \mathcal{Y}^{-}\mathcal{Y}^{+}\mathrm{Re}\Big( %
(\mathcal{Z}_{r}^{-})^2 \Big)\Big),\\
&& j^z_l=  -|\mathcal{A}|^{2} 
\Big(
 \mathcal{Z}_{l}^{-} \mathcal{Z}_{l}^{+}\Big)
\Big[(\mathcal{Y}^{-})^{2}- (\mathcal{Y}^{+})^{2}\Big]   , \qquad 
j^z_r= |\mathcal{A}|^{2} 
\Big(\mathcal{Z}_{r}^{-} \mathcal{Z}_{r}^{+}
\Big)
\Big[(\mathcal{Y}^{-})^{2}- (\mathcal{Y}^{+})^{2}\Big] 
\end{eqnarray}
and look  separately for $\partial_y j^y_{l,r} + \partial_z j^z_{l,r}$. 

We need to calculate
$\partial_y(\mathcal{Y}^{-} \mathcal{Y}^{+})$ and $ 
\partial_z (\mathcal{Z}_{r,l}^{-} \mathcal{Z}_{r,l}^{+}) $. We proceed as before replacing the derivatives by a 
judicious choice of the corresponding intertwining operator. We have
\begin{eqnarray}
\partial_y(\mathcal{Y}^{-} \mathcal{Y}^{+}) &&= 
\mathcal{Y}^{+} (\partial_y \mathcal{Y}^{-}) + 
\mathcal{Y}^{-} \partial_y(\mathcal{Y}^{+}) =\mathcal{Y}^{+} ({\hat L}^-_A- W(y))\mathcal{Y}^{-} + \mathcal{Y}^{-} (-{\hat L}^+_A +W(y)) \mathcal{Y}^{+} \nonumber \\
&&=\sqrt{E_A}\Big( (\mathcal{Y}^{+})^2- (\mathcal{Y}^{-})^2 \Big)
\label{DERY}]
\end{eqnarray}
\begin{eqnarray}
\partial_z({\cal Z}^+ {\cal Z}^-)&&=
{\cal Z}^+ \partial_z{\cal Z}^- +  {\cal Z}^- \partial_z{\cal Z}^+ ={\cal Z}^+i ({\hat L}^+_\phi -U(z)){\cal Z}^- -  {\cal Z}^- i({\hat L}^-_\phi -U(z)) {\cal Z}^+ \nonumber \\
&&=-\sqrt{E_A}\Big(({\cal Z}^+)^2 + ({\cal Z}^-)^2   \Big) ,
\label{DERZ}
\end{eqnarray}
where  the later relation is valid for both ${\cal Z}_{l,r}$. Now, let us put together the divergence  of the left current  $j_l^\mu$, for example. Substituting the required derivatives from (\ref{DERY}) and (\ref{DERZ}) we write
\begin{eqnarray}
\partial_\mu j^\mu_l=\partial_y j^y_l+  \partial_z j^z_l &&= |\mathcal{A}|^{2} \Big(
 2 \,  \mathrm{Re}\Big( (\mathcal{Z}_{l}^{+})^2 \Big) 
 \partial_y(\mathcal{Y}^{-}\mathcal{Y}^{+})
 + 
 \Big((\mathcal{Y}^{+})^{2}- (\mathcal{Y}^{-})^{2}\Big) \partial_z\Big(
 \mathcal{Z}_{l}^{-} \mathcal{Z}_{l}^{+}\Big)\Big) \nonumber \\
 &&= |\mathcal{A}|^{2} \sqrt{E_A}  \Big((\mathcal{Y}^{+})^{2}- (\mathcal{Y}^{-})^{2}\Big)\Big(
 2 \,  \mathrm{Re}\Big( (\mathcal{Z}_{l}^{+})^2 \Big) 
 - 
({\cal Z}_l^+)^2 - ({\cal Z}_l^-)^2 \Big) =0.
\end{eqnarray}
The final null result  is obtained since
\beq
\mathrm{Re}\Big( (\mathcal{Z}_{l}^{+})^2 \Big)= \frac{1}{2}\Big( (\mathcal{Z}_{l}^{+})^2 + [(\mathcal{Z}_{l}^{+})^2]^*  \Big)=  \frac{1}{2}\Big( (\mathcal{Z}_{l}^{+})^2 + (\mathcal{Z}_{l}^{-})^2]  \Big),
\eeq
with $[{\cal Z}^+_l]^*= {\cal Z}^-_l$.
A similar procedure shows that the right current  is also conserved. This yields the simultaneous conservation of the probability current $j^\mu$ and the axial current $j_5^\mu$.

}

\section{Intertwining relations}
\label{INTERT}
{The relations   (\ref{IntertElec}) are obtained by substituting the fermion components in Eqs. (\ref{SPINORS}) into the corresponding mixing equations (\ref{LREL}), and (\ref{RREL})  and subsequently using the intertwining relations 
\begin{eqnarray}
    && \hat{L}_{\phi}^- \,{\mathcal{Z}}^+(\Tilde{z})\ = s_-\sqrt{\varepsilon_A}\, {\mathcal{Z}}^-(\Tilde{z}),\qquad \hat{L}_{\phi}^+ {\mathcal{Z}}^-(\Tilde{z})\ = s_+\sqrt{\varepsilon_A}\, {\mathcal{Z}}^+(\Tilde{z}), \\
&&
    \hat{L}_A^+ \, {\mathcal{Y}}^+(y) = \sqrt{\varepsilon_A} \, {\mathcal{Y}}^-(y),\qquad
    \hat{L}_A^- \, {\mathcal{Y}}^-(y) = \sqrt{\varepsilon_A} \,{\mathcal{Y}}^+(y), 
\end{eqnarray}
with $s_+,s_-$  to be determined.
 For example, from the second equation in (\ref{LREL}) we have,  in successive steps
\begin{eqnarray}
        &&\hat{L}_{\phi}^- \psi_4(y,z) + \hat{L}_{A}^- \psi_3(y,z) = \hat{L}_{\phi}^- \left( {\mathcal{Y}}^+(y)\ {\cal D} \ {\mathcal{Z}}^+(\Tilde{z}) \right) + \hat{L}_{A}^- \left( {\mathcal{Y}}^-(y)\ {\cal C} \ {\mathcal{Z}}^-(\Tilde{z}) \right) =0, \nonumber \\          
        && \Rightarrow {\cal D}s_-\ {\mathcal{Y}}^+(y)\ {\mathcal{Z}}^-(\Tilde{z}) + {\cal C} {\mathcal{Z}}^-(\Tilde{z})   {\mathcal{Y}}^+ (y) =  \left[{\mathcal{Y}}^+(y){\mathcal{Z}}^-(\Tilde{z})\right](s_- {\cal D} + {\cal C}) =0, 
        \nonumber\\
    && \Rightarrow   s_-{\cal D}+ {\cal C}=0.
    \label{norm2} 
\end{eqnarray} 
Proceeding in a similar way, four equations for $s_{\pm}$ are found. They can be written in matrix form as
\begin{eqnarray}
    \begin{pmatrix}
        0 & 0 & s_+ & 1\\
        0 & 0 & 1 & s_-\\
        s_- & -1 & 0 & 0\\
        -1 & s_+ & 0 & 0\\
    \end{pmatrix}
    \begin{pmatrix}
        \mathcal{A}\\
        \mathcal{B}\\
        \mathcal{C}\\
        \mathcal{D}
    \end{pmatrix}=
    \begin{pmatrix}
        0\\
        0\\
        0\\
        0
    \end{pmatrix}.
\end{eqnarray}
Demanding zero determinant we obtain $s_+ \, s_- =1$, which is consistent with the general condition obtained from the factorization property. Then
\begin{eqnarray}
    \mathcal{C}=-s_-\mathcal{D}\qquad \text{and }\qquad \mathcal{A}=s_+\mathcal{B}.
\end{eqnarray}
In order to satisfy the requirement  (\ref{FINALCOND}), we need $s_-/s_+=-1$, which yields $s_-^2= s_+^2=-1$. {To fulfill these conditions we define 
\beq
s_+= i \,  \kappa= - s_-, \quad \kappa=\pm 1,
\eeq
and find that the intertwining relations in the $z$ sector introduce the additional label $\kappa$ to the spinor wave function. Throughout this work, and in all figures presented we have considered only the case with $\eta=\kappa=+1$ }.
 It is also worth remembering that the discrete spectrum of $\hat{H} _{\phi} ^{-}$ comes from the relation (\ref{CONDE}) [$E_A = -E_{\phi}$] and that the eigenfunctions $\{ \mathcal{Z} ^{\pm} _{n} (\Tilde{z}) \}$ satisfy the intertwining relations
\begin{align}\label{IntertElec}
\mathcal{Z} ^{\pm} _{n} (\Tilde{z}) = \mp \frac{i}{\sqrt{ E _{\phi \, n} ^{+} }} \hat{L} _{\phi} ^{\pm} \, \mathcal{Z} ^{\mp} _{n} (\Tilde{z})  , \qquad n =1,2,3, \cdots .
\end{align}

\bibliography{bib}

\end{document}